# How does stock market reflect the change in economic demand?
# A study on the industry-specific volatility spillover networks of China's stock market during the outbreak of COVID-19

Fu Qiao[*]   Yan Yan[*]

**Abstract**：Using the carefully selected industry classification standard, we divide 102 industry securities indices in China's stock market into four demand-oriented sector groups and identify demand-oriented industry-specific volatility spillover networks. The "demand-oriented" is a new idea of reconstructing the structure of the networks considering the relationship between industry sectors and the economic demand their outputs meeting. Networks with the new structure help us improve the understanding of the economic demand change, especially when the macroeconomic is dramatically influenced by exogenous shocks like the outbreak of COVID-19. At the beginning of the outbreak of COVID-19, in China's stock market, spillover effects from industry indices of sectors meeting the investment demand to those meeting the consumption demands rose significantly. However, these spillover effects fell after the outbreak containment in China appeared to be effective. Besides, some services sectors including utility, transportation and information services have played increasingly important roles in the networks of industry-specific volatility spillovers as of the COVID-19 out broke. By implication, firstly, being led by Chinese government, the COVID-19 is successfully contained and the work resumption is organized with a high efficiency in China. The risk of the investment demand therefore was controlled and eliminated relatively fast. Secondly, the intensive using of non-pharmaceutical interventions (NPIs) led to supply restriction in services in China. It will still be a potential threat for the Chinese economic recovery in the next stage.

[*] University of Chinese Academy of Sciences, School of Economics and Management; Contact author: qiaofu17@mails.ucas.edu.cn

## 1. Introduction

In stock market, it is a frequent occurrence that the fluctuations in stock prices initially occur in companies belonging to one industry sector, and gradually spread to those belonging to other ones. China's stock market has become the second largest stock market all over the world. Up to January 2020, 3780 companies belonging to variety of sectors listed their shares in China's stock market. The market value of these shares was more than 60.38 billion RMBs (roughly equals to 8.65 trillion US dollars). Thus, it is fairly important for both investors and policymakers all over the world to understand the complex linkage effect showed in fluctuations in stock prices (or yields) of companies of different sectors in China's stock market.

One of the typical measurements of the linkage characteristics between different variables is spillover effect. The spillover effect mainly consists of the mean spillover effect and the volatility spillover effect. This paper highlighted the volatility spillover effect, which reflects the contagion direction of information and the transmission structure of risk between different securities. Theoretically, the volatility spillover effect is led by the incomplete information. Any occurrence of the new information will possibly lead to price changes in multiple classes of securities. However, it takes time for information being transmitted between the investors with preferences for different classes of securities. Investors focusing on a specific securities class will always obtain the information that possibly has direct influence on the prices of securities of this class earlier. The investors' out-of-sync information observing leads to the out-of-sync price changes in securities of different classes. As the result, the volatility spillover effect occurs among the fluctuations in different securities prices [1]. The linkage level of volatility spillover effect can be measured by using GARCH family models (models such as BEKK-GARCH and DCC-GARCH) or the variance decomposition model under the VAR framework. [2-5]. Research on spillover effect across markets used to employ daily data, while high-frequency data has been increasingly used in recent literature [6-8].

It has become a popular perspective analyzing the interaction across social and economic units basing on the concepts of network topology [9]. A significant strand of literature has introduced the theory of complex networks into the study when analyzing the spillover effect across multiple financial institutions or markets [10, 11]. The literatures on spillover networks analysis of financial markets mainly focused on the spillovers across financial markets in different countries' (or regions') [12, 13], the spillovers between different asset categories and those between different industry sectors [14-16].

For researches on spillover networks analysis across financial markets of different countries', there have been conclusions with high consistency and interpretability in. According to existing literatures on this topic, the nodes representing the securities indices or portfolios of financial markets of developed countries always played a dominated role in the spillover networks [17, 18]. Many researches also illustrated that the spillovers between the financial markets of emerging market countries and those of developed ones have rapidly increased after the financial crisis in 2007, while they were relatively isolated with each other before the financial crisis [19-21].

Owing to two main reasons, the findings of researches on networks of spillovers across global financial markets are inspiring and easy to understand. On the one hand, the nodes representing financial markets of different countries or regions' can be easily classified with geographic or economic standards [13, 22]. On the other hand, there are various extensively researched theories of international economy and finance, which can be introduced to explain the spillovers.

Take another look at researches on industry-specific spillover networks analysis. Early researches on this topic showed that exogenous shocks to macro economy of one country do not lead to the fluctuations in prices of all securities in this country at the same time [23-25]. Being inspired by these literatures, we further considered demand structure influences industry-specific spillover networks. The demand of a country mainly consists of the demand for consumption, investment and export. One of the critical things to the profit and asset price of the companies that if their goods or services successfully meet a part of demand or not [26, 27]. The exogenous shock has different potential influences on different parts of demand, and it will change the companies' profitability and their asset prices at different levels depending on the industry sector they belong. The securities industry index can be viewed as a representative of the asset price of all companies in this industry sector. The influence of exogenous shock on economic demand therefore should be reflected in the structural change in the industry-specific volatility spillover networks.

Differing from the early researches, the recent ones on industry-specific volatility spillover networks highlighted the measurement of the linkage level. These researches were initially motivated by finding the arbitrage opportunities between assets of upstream industry sectors and those of downstream ones in supply chain. Besides, literatures like [12, 28] further found that the volatility spillovers also exists between industry sectors without direct input-output relationship.

However, existing literatures did not answer the question about how could industry-specific volatility spillover networks reflect the economic demand and its changes. We believe that there are two reasons should be blamed. Firstly, it is more difficult to give a proper explanation for the findings in industry-specific volatility spillover networks analysis than giving explanation for those of analysis across countries or regions. A part of the detected spillovers in the networks might match the economics theory, such as the spillovers between the energy sector and the finance sector, or those of the transportation sector and the consumption sector [16, 29]. However, the rest of spillovers in networks might not be properly explained. Especially, when introducing the max spanning tree or some clustering methods, spillover paths or sub-networks that consist of a large number of nodes will be generated. Then it would be even harder to explain for the findings [30]. Secondly, some scholars have criticized the arbitrariness when selecting industry classification standards. Literature [31] pointed out that the industry classification standard should be cautiously selected depending on the research targets. When it is necessary, self-built industry indices should also be used for pursuing more meaningful numerical results and theoretical implication [31, 32].

Most of the researches on industry-specific spillover network analysis of China's stock market chose the industry classification standard made by China Securities Index Co., ltd (the CSI standard) [12,28,30]. However, the CSI standard cannot match all research targets. According to the CSI standard, the categories of "consumption goods" and "capital goods" are not parallel with each other. Companies supplying goods or services for meeting the consumption demand can be classified as sectors (level 1 categories) "consumer staples" or "consumer discretionary". In contrast, companies meeting the investment demand cannot be classified as a sector. They can only be classified as an industry group (a level 2 category) "capital goods" which belongs to the sector "industrial". Thus, the spillover network analysis using the CSI standard cannot reflect the economic demand and its change in a proper way.

Inspired by the structural transformation theory, we believe that the industry sectors can be further classified as groups by the kind of the demand their goods or services meet. Literatures [33] and [34] used the World Input-Output Tables (WIOT, [35]) to reorganize the industry classification from the perspective of the demand. Note that the WIOT is not suitable for classifying listed companies. Instead, we introduced the industry classification standard made by SWS Research Co., Ltd., which is the largest securities research institute in Mainland China. The SWS standard divided the industry sectors into four sector groups, each of which is homogenous in meeting specific economic demand. Specifically, the "equipment manufacturing" and "capital goods" are sector groups meeting the investment demand in China. The "consumption goods" is the group meeting the consumption demand. Besides, some service sectors meet neither purely the consumption demand nor the investment demand. Those sectors are therefore classified as a group called "unclassified services". We call the industry-specific volatility spillover network basing on the SWS standard as the demand-oriented industry-specific volatility spillover network. This network can reflect the economic demand change better through the financial market than the network basing on CSI standard.

Therefore, using the SWS standard, we identified the GARCH-BEKK based demand-oriented industry-specific volatility spillover networks of China's stock market. Each node in the networks represents a level 2 industry securities index in the SWS industry list. We chose the minute-per-minute return data of 102 SWS industry indices as the sample. The study period is between the January 2 and March 20, 2020. In this period, as a typical exogenous shock, the outbreak of the COVID-19 dramatically changed the economic demand in China.

Recent literatures have reported the influence of COVID-19 on both the macro economy and financial markets of different countries or regions'. Some of them focused on the impact of the disease on the financial market in a single country [36-38] or the overall impact on global financial markets [39, 40]. Furthermore, in literatures [41, 42], industry-specific networks were identified basing on macroeconomic data rather than the data from financial markets. These literatures provided us good inspirations for further designing the research on illustrating how industry-specific volatility spillover networks can reflect the economic demand change. Our study extends the literature and contributes to the following aspects.

(1) From the perspective of demand, we purposed a new idea for reconstructing the structure of the industry-specific spillover network. By reorganizing the industry securities indices into demand-oriented sector groups, a better linkage between the theories of macroeconomics and the industry-specific network analysis of financial market can be therefore obtained.

(2) We provided an early report of the structural change in the industry-specific volatility spillover networks of China's financial market around the outbreak of COVID-19. We further analyzed how the changes in this network reflected the changes in economic demand caused by the disease. Meanwhile, because of using the intraday data, we avoided being trapped by the "dimension curse" [43] which is common in research on short and rapid exogenous shocks.

(3) A list of new economic implications was found from the numerical result. Firstly, during the entire study period, there were stable spillovers from the sector group "capital goods" to the group "consumption goods". The spillovers from sector groups "capital goods" and "equipment manufacturing", which are represent the demand for investment, to other sector groups rose significantly at the beginning of the outbreak of COVID-19. However, these spillovers fell about one month later. Secondly, the level of spillovers from the sector group "unclassified services" was continuously rising during the whole study period. This rising trend reflected that the intensive using of non-pharmaceutical interventions (a.k.a. NPIs) [44] in China caused the supply restriction in services, and therefore had an overall impact on all kinds of demand.[44] Researches using the CSI standard usually found the consumer discretionary sector or the industrial sector as the origin of volatility spillovers of the networks [12, 45].[45] Our findings are consistent with existing literatures to many aspects. However, considering the extreme exogenous shock like the outbreak of COVID-19, this paper provided a better way explaining how financial market reflects the demand changes, especially the significant and long-lasting influence of services sectors which is easily neglected by other researches.

The next section introduces the data selection and pre-processing. Section 3 discusses the methodology. Section 4 presents the empirical study of the demand-oriented industry-specific volatility spillover network analysis based on the SWS industry classification standard. Section 5 is the further discussion, and section 6 concludes.

## 2. Data

### 2.1 The study period

The study period is between January 3 and March 20, 2020. As early as December 27, 2019, the local government of Wuhan began to report the patients of the pneumonia, which was unidentifiable at that time, and took public health responses to the infection. As of January 20, 2020, Chinese government began to implement nationwide containment of the COVID-19. On January 31, 2020, the World Health Organization (WHO) declared the COVID-19 a public health emergency of international concern (PHEIC). To guarantee that all patients can be treated, Chinese government covered all bills of the pharmaceutical treatment with the budgets. Besides, to reduce the size of the pandemic, multiple NPIs were used by Chinese government as well including the inter-city travel restrictions, the early identification and isolation of the suspected ill persons and the contact restrictions measures [44]. As the result, the outbreak was preliminarily contained in China at the end of February. Since March 18, 2020, the number of new patients were maintaining lower than 10 per day. However, COVID-19 began to breakout outside China. On February 29, 2020, the WHO increased the assessment of the risk of spread to "very high" at global level.

Taking into account the size of spread and the progress of containing of the COVID-19 both inside and outside China, we divided the study period into three sub-periods. The period 1 is between January 3 and January 23, 2020. The period 2 is between February 3 and February 28, 2020. The period 3 is between March 2 and March 20. Period 1, 2 and 3 has 16, 20 and 15 trading days, respectively.

### 2.2 The industry classification standard

Industry classification is a long-standing problem in financial research [32]. There are two kinds of industry classification standards, the administrating-oriented standards and the investing-oriented ones. Administrating-oriented standards define the categories mainly according to the companies' production technology. This kind of standards is issued by governments or market supervisors for purpose of supervising the companies by industry categories. In contrast, investing-oriented standards not only pay attention to the production technology, but also highlight the kind of market demand met by the company. This kind of standards is made to meet the need of the investment professionals.

The typical administrating-oriented standards consist of the International Standard Industrial Classification of All Economic Activities (ISIC) and North American Industry Classification System (NAICS). The popular investing-oriented ones consist of the Global Industry Classification Standard (GICS) and FTSE Global Classification System. In China's financial market, the China Securities Regulatory Commission issued an administrating-oriented standard called the Guideline for Industry Classification of China's Listed Companies in 2012. Besides, Chinese Securities Index Co., Ltd. and SWS Securities Co., Ltd established two different investment-oriented industry classification standards, which are the CIS standard and SWS standard, respectively.

The industry securities indices traded in the exchanges are always made according to one of the investing-oriented standards. Among the investing-oriented standards used in China' financial market, the CSI standard is the one closest with the GICS standard. This standard is therefore the most popular one chosen for the spillover networks analysis of China's stock market. However, the categories representing consumption and investment are unbalanced according to the CSI standard. Consuming goods firms are classified as two sectors (level 1 categories), consumer stable and consumer discretionary, while capital goods ones is only an industry group (level 2 category). Thus, we chose the SWS standard as the industry classification standard. According to SWS standard, the securities in China's stock market are divided into 28 sectors, which are further divided into 104 industry groups. As shown in Table 1, the most significant difference from the CSI standard is that SWS standard integrated the sectors into four demand-oriented sector groups.

**Table 1: Official Categories of sector groups and sectors according to SWS industry classification standard**

| Sector Group (Abbreviation) | Sector | The last 4 digits of the codes of relative industry group indexes (level 2 categories of industry classification system) |
|---|---|---|
| Consumption goods (Cg) | Agriculture, forestry, husbandry and fishery | 1011,1012,1013,1014 1015,1016,1017,1018 |
| | Household appliances | 1111,1112 |
| | Food and beverage | 1123, 1124 |
| | Apparel and textiles | 1131,1132 |
| | Light manufacturing | 1141,1142,1143 |
| | Biochemical and pharmaceuticals | 1151,1152,1153,1154,1155,1156 |
| | Leisure Services | 1211,1212,1213,1214 |
| | Commercial trade | 1202,1203,1204,1205 |
| Capital goods (Kg) | Mining | 1021,1022,1023,1024 |
| | Chemicals | 1032,1033,1034,1035,1036,1037 |
| | Non- ferrous metal | 1051,1053,1054,1055 |
| | Construction and decorations | 1711,1712,1713 |

|  | Building materials | 1721,1722,1723,1724,1725 |
|  | Ferrous metal | 1041 |
| Equipment manufacturing (Ke) | Machinery | 1072,1073,1074,1075,1076 |
|  | Electronic components | 1081,1082,1083,1084,1085 |
|  | Electrical equipment | 1731,1732,1733,1734 |
|  | Motor | 1092,1093,1094,1881 |
|  | Defense and military industry | 1741,1742,1743,1744 |
|  | Information facilities | 1222,1223 |
| Unclassified services (Us) | Utilities | 1161,1162,1163,1164 |
|  | Transporting | 1171,1172,1173,1174, 1175,1176,1177,1178 |
|  | Real estate | 1181,1182 |
|  | Bank | 1192 |
|  | Non-bank financial services | 1191,1193,1194 |
|  | Information services | 1222,1223 |
|  | Media | 1751,1752,1761 |

Note: 1. The conglomerates sector consists of listed companies with diversified businesses in which no single business is dominant. Although as a sector vertex in the sector-specific spillover network, SWI conglomerates sector index is belong to none of the sector groups.

2. The codes of industry group indexes (level 2 categories of industry classification system) consists of 6 digits in which the first 2 digits are "80-"

Table source: Wind Financial database

As a supplement of Table 1, we listed the names and the codes of all industry group indices according to SWS standard. In addition, because the level 3 categories (industry) was not mentioned in this paper, we will call the "industry group securities index" as "industry securities index" for short in the following sections.

**2.3 The data pre-processing and descriptive statistics**

The minute-per-minute data of the close prices of 102 SWS industry securities indices are available in the Wind Financial Database. By convention, we calculate the log-return of the $P_{i,t}$, which is the price of the index i at the moment t, as $r_{i,t} = ln(P_{i,t}) - ln(P_{i,t-1})$. Then we obtain $r_i = \{r_{i,t}\}, t \in \{1,2,\cdots,T\}$ as the log-return series of the index i. The comprehensive descriptive statistics the log-return series of all industry securities indices in different periods can be found in appendix. In the body of text, we only displayed the distributions of the means and variances of the indices' log-returns in different period in Figure 1.

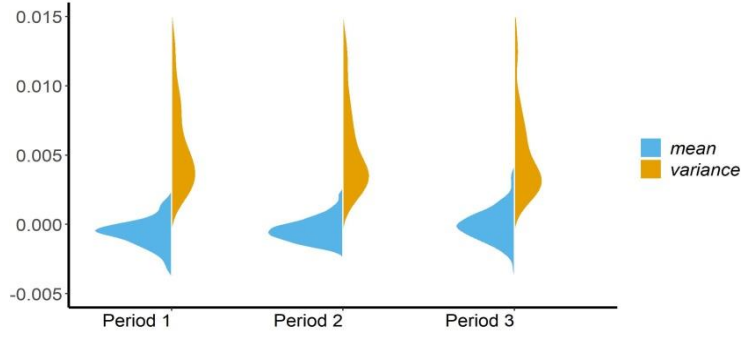

**Figure 1**：The distributions of means and variances of the minute-per-minute return of 102 SWS industry indexes in different periods

### 3. Methodology
#### 3.1 The GARCH-BEKK based volatility spillover network

GARCH-BEKK model is purposed by [46] [46]. The economic implication of the model is attracting because its parameters are able to detect the spillover effect between the variables. Consider time series $r_i = \{r_{i,t}\}, t \in \{1,2,\cdots,T\}$ and $r_j = \{r_{j,t}\}, t \in \{1,2,\cdots,T\}$, to test the spillover effect between them, a bi-variable GARCH-BEKK model is required. The bi-variable GARCH-BEKK model consists of a mean equation and a variance equation. According to literature [47, 48] [47][48], the lag order of both equations can be set as 1. The mean equation of bi-variable GARCH-BEKK model is showed in equation (1):

$$\begin{bmatrix} r_{i,t} \\ r_{j,t} \end{bmatrix} = \begin{bmatrix} \mu_i \\ \mu_j \end{bmatrix} + \begin{bmatrix} \varphi_{ii} & \varphi_{ij} \\ \varphi_{ji} & \varphi_{jj} \end{bmatrix} \begin{bmatrix} r_{i,t-1} \\ r_{j,t-1} \end{bmatrix} + \begin{bmatrix} \varepsilon_{i,t} \\ \varepsilon_{j,t} \end{bmatrix} \quad (1)$$

where $\mu_i, \mu_j, \varphi_{ii}, \varphi_{ij}, \varphi_{ji}$ and $\varphi_{jj}$ are parameters to be estimated, the $\varepsilon_{i,t}$ and $\varepsilon_{j,t}$ are residuals. They are also called innovations, which represent the influence of the new information generated at moment $t$.

The variance model is as equation (2):

$$H_t = C'C + A'\varepsilon_{t-1}\varepsilon'_{t-1}A + B'H_{t-1}B \quad (2)$$

where $H_t = \begin{bmatrix} H_{ii,t} & H_{ij,t} \\ H_{ji,t} & H_{jj,t} \end{bmatrix}$ represents the conditional covariance matrix of $r_i$ and $r_j$, $\varepsilon_t = \begin{bmatrix} \varepsilon_{i,t} \\ \varepsilon_{j,t} \end{bmatrix}$ is the innovation vector. $C = \begin{bmatrix} C_{ii} & 0 \\ C_{ij} & C_{jj} \end{bmatrix}$, $A = \begin{bmatrix} a_{ii} & a_{ij} \\ a_{ji} & a_{jj} \end{bmatrix}$ and $B = \begin{bmatrix} b_{ii} & b_{ij} \\ b_{ji} & b_{jj} \end{bmatrix}$ are parameters.

To detect the volatility spillovers between $r_i$ and $r_j$, following hypothesizes are to be tested.

$$H_0: a_{ij} = a_{ji} = b_{ij} = b_{ji} = 0 \quad (3)$$
$$H_1: a_{ij} \neq 0 \text{ or } a_{ji} \neq 0 \text{ or } b_{ij} \neq 0 \text{ or } b_{ji} \neq 0 \quad (4)$$

in which $H_0$ is the null hypothesis and $H_1$ is the alternative hypothesis. By convention, we reject null hypothesis at 90% confidence level. If the null hypothesis is rejected, it means that there are spillovers between $r_i$ and $r_j$. Specifically, the direction of spillover effect is from $r_i$ to $r_j$ when $a_{ij} \neq 0$ or $b_{ij} \neq 0$. Otherwise, the direction is from $r_j$ to $r_i$ when $a_{ji} \neq 0$ or $b_{ji} \neq 0$.

To test volatility spillovers between multiple variables, a set of bi-variable GARCH-BEKK models are required. After all testing finished, we can identify the GARCH-BEKK based volatility spillover networks. Let $Net(V,E)$ represents the industry-specific securities indices volatility spillover networks. The set $V = \{v_1, v_2, \cdots, v_N\}$ represents the vertices, which are called nodes as well, of industry securities indices. Each of the nodes $v_i$ is characterized by a log-return time series $r_i$. The set $E$ represents the edges of networks. For $\forall v_i, v_j \in V$, the edges from $v_i$ to $v_j$ satisfy the indicator function $e_{ij}$ as follows:

$$e_{ij} = \begin{cases} 1 & if\ i \neq j\ and\ there\ is\ volatility\ spillover\ from\ r_i\ to\ r_j \\ 0 & otherwise \end{cases} \quad (5)$$

Now consider the weight of $e_{ij}$. By reference to literatures [49, 50] [49,50], we calculated the weight of edges as follows

$$w_{ij} = |a_{ij}| + |b_{ij}| \quad (6)$$
$$w_{ji} = |a_{ji}| + |b_{ji}| \quad (7)$$

where the weights of $e_{ij}$ and $e_{ji}$ are represented $w_{ij}$ and $w_{ji}$, respectively. Now the weighted and directed GARCH-BEKK based volatility spillover networks have been identified. The intensity of the edge $e_{ij}$ can be calculated as $s_{ij} = e_{ij} w_{ij}$

### 3.2 The nodes' importance ranking indicators
#### 3.2.1 The indicators at node level
1. Connectivity and relative influence

The connectivity indicators consist of the $O_i$ and $I_i$, which represent the total intensity of outward spillovers from $v_i$ and the total intensity of inward spillovers to $v_i$, respectively. $O_i$ and $I_i$ are calculated as:

$$O_i = \sum_{j=1, j\neq i}^{N} s_{ij} \quad (8) \qquad I_i = \sum_{j=1, j\neq i}^{N} s_{ji} \quad (9)$$

Both $O_i$ and $I_i$ are absolute indicators. The relative influence of $v_i$ is calculated as:

$$ri_i = \begin{cases} \dfrac{O_i - I_i}{O_i + I_i} & , otherwise \\ 0 & , if\ O_i + I_i = 0 \end{cases} \quad (10)$$

2. Weighted k-shell decomposition

Besides the number of neighbors, the location of a node in the network is also critical for the assessment of its importance. Literature [51] therefore purposed the k-shell decomposition to evaluate the locational importance of the nodes. [51] The k-shell decomposition is the method that reshapes the networks into a layered structure according to their connectivity patterns. For an un-weighted network $N_0 = Net(V_0, E_0)$, the layer $L$ of $N_0$ is a subset of nodes, each of which has only one neighbor. We assign the layer an integer label $L_1$ and remove it from the $N_0$. Then we obtain a new network $N_1 = Net(V_1, E_1)$. Similarly, we identify the layer of the network $N_{n-1}$, assign the layer a label $L_n$ and remove it from the network. After repeating the step for $K$ times, each of the nodes in the original network $N_0$ can be assigned to one of the layers. The k-shell decomposition is a coarse-grained ranking method. The nodes belonging to the layer with label $L_1$ are those with the lowest locational importance in the networks. Relatively, the nodes belonging to the layer $L_K$ are those with the highest locational importance. in another word, they are the center of the network.

The vanilla k-shell decomposition fails to consider the intensity of connections. Thus, it cannot rank the nodes for weighted networks. The literature [52] [52]extended the vanilla k-shell decomposition to weighted k-shell decomposition. An alternative measure for node degree is purposed in [52], which is shown in equation (11).

$$wk_i = [k_i^\alpha O_i^\beta]^{\frac{1}{\alpha+\beta}} \quad (11)$$

where $k_i$ represents the number of neighbors connected with $v_i$, $O_i$ is defined in equation (8). According to [52], the parameters can be set as $\alpha = \beta = 1$. Note that $wk_i$ is not integer value. Therefore, we should firstly divide all the weights of edges in $N_0$ with their minimum value, and discretize the resulting weights by rounding to their closest integer. Then we get $N_0' = Net(V_0, E_0')$, in which the minimum value of the weights equals to 1. Each step of the weighted k-shell decomposition consists of: first, normalize $N_n$ to $N_n'$, second, identify the layer $k = n + 1$ and third, remove the layer from $N_n$ and obtain the $N_{n+1}$.

3. Betweenness centrality

The betweenness centrality is based on the shortest distances between the nodes. For weighted directed networks, the definition of the shortest distance from $v_i$ to $v_j$ is shown in equation (12) [53]

$$d_{ij}^\alpha = \min_{D_{ij}}\{(s_{id_1}^{-\alpha} + s_{d_1 d_2}^{-\alpha} + \cdots + s_{d_k j}^{-\alpha}), s_{ij}^{-\alpha}\} \quad (12)$$

where $D_{ij} = \{v_{d_1}, v_{d_2}, \cdots, v_{d_k}\}, 1 \leq k \leq N - 2$ represents a set of arbitrary intermediate nodes of the spillover paths from $v_i$ to $v_j$. The set of the intermediate nodes of the shortest path can be therefore defined as $D_{ij}^*$. If $d_{ij}^\alpha = s_{ij}^{-\alpha}$, there are no intermediate nodes in the shortest path from $v_i$ to $v_j$, and $D_{ij}^* = \emptyset$ [53]. For unweighted networks, the parameter $\alpha$ simply equals to zero. For weighted networks, the value of $\alpha$ depends on the relationship between the link intensity between the nodes and their distance. In the weighted and directed spillover networks of financial markets, firstly, the higher link intensity between the nodes means the shorter distance. The value of $\alpha$ should be positive. Secondly, if there are more intermediate nodes in the spillover path between two nodes, their distance is longer. The $\alpha$ is supposed to be less than 1. With a comprehensive consideration, we take $\alpha = 0.5$.

We can easily define the betweenness and the closeness based on the definition of the shortest distance $d_{ij}^\alpha$. The weighted betweenness centrality of $v_i$, $WBC_i$, represents the proportion of shortest paths from $v_j$ to $v_k$, which include $v_i$ as an intermediate node. The $WBC_i$ can be calculated by equation (13).

$$WBC_i = \sum_{i \neq j \neq k} \frac{g_{jk}(i)}{g_{jk}} \quad (13)$$

where $g_{jk}$ is the number of different $D_{jk}^*$, $g_{jk}(i)$ is the number of $D_{jk}^*$ including $v_i$ as an intermediate node.

**3.2.2 The indicators at group level**

When dividing the nodes into several groups, additional indictors are required for assessing the importance of the groups according to [10, 54][54]

1. Group-specific connectivity

The outward spillover effect from the noes in one group to those in other groups can be defined as "total out to other" (TOTO). Similarly, the inward spillover to one group from other groups can be defined as "total in from other" (TIFO). The TOTO and TIFO are showed in equation (14) and equation (15):

$$TOTO_i = \sum_{j=1}^{N-N_m} s_{ij}, v_j \in V \setminus V_m \quad (14)$$

$$TIFO_i = \sum_{j=1}^{N-N_m} s_{ji}, v_j \in V \setminus V_m \quad (15)$$

where $v_i$ belongs to the subset $V_m = \{v_i | i \in m\}$ which includes $N_m$ nodes.

2. Sector influence

Let $V_m, V_n \subset V$ which satisfies $V_m = \{v_i | i \in m\}$ and $V_n = \{v_j | j \in n\}$. We calculate the sector influence indicator as follows:

$$SI_{mn} = \frac{1}{N_m N_n} \sum_{i \in m} \sum_{j \in n} s_{ij} \quad (16)$$

The higher the $SI_{mn}$ is, the more intensive the spillover is from the subset $V_m$ to the subset $V_n$.

**3.3 Earth mover's distance (EMD)**

None of the indicators introduced in section 2.2 highlight measuring the distributional change in the spillover intensity. Therefore, we introduce the earth mover's distance (EMD) to consider the intensity distributions change in spillovers across different groups. The EMD [55] is a cross-bin distance that is defined as the minimal cost that must be paid to transform one histogram into the other. An intensity distribution of spillovers can be represented by countable clusters. Each cluster is represented by its mean, and by the fraction of the distribution that belongs to that cluster. We call such a representation the signature of the distribution. Then the distributional change in spillovers intensity between period 1 and period 2 can be formalized and solved as a transportation problem. We transform the distribution of link strength in period 1 and period 2 into signatures $S_1$ and $S_2$.

$$\begin{cases} S_1 = \{(S_{11}, p_{S_{11}}), (S_{12}, p_{S_{12}}), \cdots, (S_{1m}, p_{S_{1m}})\} \\ S_2 = \{(S_{21}, p_{S_{21}}), (S_{22}, p_{S_{22}}), \cdots, (S_{2n}, p_{S_{2n}})\} \end{cases} \quad (17)$$

where the intensity distribution of the spillovers in period 1 and period 2 are discretized into $m$ and $n$ clusters. The $S_{1i}$ and $S_{2j}$ represent the means of the $ith$ cluster in period 1 and the $jth$ one in period 2, respectively. Both $S_{1i}$ and $S_{2j}$ are one dimensional real value, we define the ground distance between $ith$ cluster in period 1 and the $jth$ one in period 2 as $d_{ij} = |S_{1i} - S_{2j}|$. The $p_x$ represents the weight of cluster $x$. Besides, the $p_x$ naturally satisfies $\sum_{i=1}^{m} p_{S_{1i}} = \sum_{j=1}^{n} p_{S_{2j}} = 1$.

We want a flow $F = [f_{ij}]_{m \times n}$ with positive $f_{ij}$ from $S_{1i}$ to $S_{2j}$ that minimizes the transporting cost

$$C(P_1, P_2, F) = \sum_{i=1}^{m} \sum_{j=1}^{n} f_{ij} d_{ij} \quad (18)$$

s.t. $\quad f_{ij} \geq 0 \quad\quad 1 \leq i \leq m, 1 \leq j \leq n \quad (19)$

$$\sum_{j=1}^{n} f_{ij} = p_{S_{1i}} \quad\quad 1 \leq i \leq m \quad (20)$$

$$\sum_{i=1}^{m} f_{ij} = p_{S_{2j}} \qquad 1 \leq j \leq n \tag{21}$$

$$\sum_{i=1}^{m}\sum_{j=1}^{n} f_{ij} = \min\left(\sum_{i=1}^{m} p_{S_{1i}}, \sum_{j=1}^{n} p_{S_{2j}}\right) = 1 \tag{22}$$

Constrain (20) and (21) represent the supply and demand constrain in transportation problem. Constrain (22) represents the total flow constrain. The problem can be solved with Hungarian method.

### 4. The demand-oriented industry-specific volatility spillover networks analysis
### 4.1 The network before the CoVID-19 outbreak

Table 2 reports the summary of the nodes in period 1 by group. According to Panel A, before the breakout of COVID-19, the sector group Kg is a significant volatility supplier in the network. The median of $O_i$ and that of $TOTO_i$ of the nodes in the group Kg are 6.71 and 5.39, much higher than other group counterparts. The indicator $\sum TOTO/\sum O$ measures the proportion that outward spillovers from one sector group to accounts for in its total spillovers. The group unclassified services (Us) has the highest value of $\sum TOTO/\sum O$, which is 78.55%.

According to Panel B, the median of $I_i$ and that of $TIFO_i$ of the group equipment manufacturing (Ke) is 5.67 and 4.48. Meanwhile, the value of $\sum TIFO/\sum I$ of this group is also the highest, which is 78.58% in its total $I_i$.

Panel C showed that group Kg is the only one with a positive median of $ri_i$ (0.09). Those of other three groups are all negative values in which the lowest one is the median of $ri_i$ of the group Cg (-0.15). The median of $WBC_i$ of the group Kg is 39, which is also much higher than other group counterparts are. The median of $WBC_i$ of the group Us is 16, which is the lowest. It means that many strong and critical spillover paths go through the group Kg while few of them go through the group Us.

**Table 2: The summary for indicators of the nodes by sector groups in period 1(before the CoVID-19 outbreak)**

| Panel A: sector groups as contributors of spillovers | | | | | | | |
|---|---|---|---|---|---|---|---|
| Sector group | O | | | TOTO | | | $\frac{\sum TOTO}{\sum O}$ (%) |
| | Min | Median | Max | Min | Median | Max | |
| Ke | 1.12 | 4.50 | 11.16 | 0.98 | 3.54 | 7.08 | 73.73 |
| Cg | 0.09 | 4.54 | 20.43 | 0.05 | 2.56 | 16.20 | 70.97 |
| Kg | 1.71 | 6.71 | 16.66 | 1.49 | 5.39 | 12.10 | 76.89 |
| Us | 0.44 | 4.55 | 13.87 | 0.41 | 3.81 | 11.14 | 78.55 |
| Panel B: sector groups as receivers of spillovers | | | | | | | |
| Sector group | I | | | TIFO | | | $\frac{\sum TIFO}{\sum In}$ (%) |
| | Min | Median | Max | Min | Median | Max | |
| Ke | 3.05 | 5.67 | 9.65 | 2.34 | 4.48 | 7.54 | 78.58 |
| Cg | 1.92 | 5.21 | 11.77 | 1.40 | 3.84 | 9.04 | 73.84 |
| Kg | 2.22 | 5.50 | 8.47 | 1.66 | 3.69 | 6.42 | 69.83 |
| Us | 1.03 | 4.61 | 10.30 | 0.77 | 3.77 | 8.45 | 78.10 |
| Panel C: the net spillovers and centrality indicators | | | | | | | |
| Sector | Ri | | | | WBC | | |

| group | Min | Median | Max | Min | Median | Max |
|---|---|---|---|---|---|---|
| Ke | -0.72 | -0.09 | 0.46 | 4 | 17.5 | 104 |
| Cg | -0.95 | -0.15 | 0.51 | 0 | 21 | 230 |
| Kg | -0.52 | 0.09 | 0.70 | 0 | 39 | 183 |
| Us | -0.73 | -0.05 | 0.67 | 0 | 16 | 113 |

Note: Ke, Cg, Kg and Us are the abbreviations of the sector groups of equipment manufacturing, consumption goods, capital goods and unclassified services, both here and below.

Table 3 reported the inter-group spillovers of the network in period 1. We can find the significant asymmetry in the spillover between the group Kg and Cg. Both the gross and net spillovers from Kg to Cg are all the highest, which are 49.02 and 16.22. Besides, the net spillover from Kg to Ke is 14.8. All of the rest of inter-group spillovers are lower than 10. Relatively, there are only slight asymmetry in the spillovers between other sector groups, except that between Kg and Cg. The net spillover from Cg to Ke is much weaker than that from Kg to Cg. Thus, viewing the group Kg and Ke as integral whole sector group meeting the investment demand, it has net spillover to the group Cg.

In period 1, the group Kg is the main contributor of spillovers from all aspects. The outward spillovers from Kg to other groups account for 23.2% in the total spillovers in the network. In contrast, the Cg plays the role of main receiver of spillovers, which receives 22.3% of the total spillovers. The group Ke also receives 20.3%, a proportion only slightly lower than that of Cg, of the total spillover effects. In addition, all other three groups have net spillovers to the group Ke.

**Table3: The cross sector group analysis of volatility spillovers in period 1 (before the COVID-19 outbreak)**

| From\to | Intensity of spillovers | | | | | No. of direct spillover paths | | | | |
|---|---|---|---|---|---|---|---|---|---|---|
| | Ke | Cg | Kg | Us | Total | Ke | Cg | Kg | Us | Total |
| Ke | 30.16 | 36.73 | 26.05 | 21.04 | 537.55 | 223 | 307 | 207 | 222 | 4308 |
| Cg | 40.64 | 43.24 | 32.80 | 30.79 | | 295 | 374 | 268 | 268 | |
| Kg | 40.81 | 49.02 | 37.67 | 34.62 | | 284 | 372 | 247 | 245 | |
| Us | 27.74 | 34.50 | 27.19 | 24.54 | | 217 | 292 | 202 | 198 | |

**4.2 The network at the beginning of the CoVID-19 outbreak**

Table 4 reports the summary of the nodes in period 2 by group. According to Panel A, the median of $O_i$ and that of $TOTO_i$ of the group Kg is 8.81 and 6.39, which are even higher than those in period 1. The group Kg also has the highest value of $\sum TOTO/\sum O$, which is 76.21%. The median of $O_i$ and that of $TOTO_i$ of group Us is 6.15 and 4.46, which increase the most significantly comparing to those in period 1.

According to Panel B, the median of $I_i$ and that of $TIFO_i$ of group Ke are the highest, which are 8.81 and 6.39. The median of $I_i$ and that of $TIFO_i$ of group Cg is 5.58 and 4.65. Meanwhile, the value of $\sum TIFO/\sum I$ of this group is also the highest, which exceeds 80%.

According to Panel C, the highest median of $ri_i$ is that of group Kg, which is 0.2. The lowest one is that of the group Cg, which is -0.31. In period 2, the difference of $ri_i$ among sector groups enlarge compared with that in period 1. The median of $WBC_i$ of the group Kg is 33, which is still the highest. However, the median of $WBC_i$ of the group Us is 27, which increases rapidly. It means the centrality of nodes in the group Us become much higher.

**Table 4: The summary for indicators of the nodes by sector groups in period 2 (at the beginning of the CoVID-19 outbreak)**

| Panel A: sector groups as givers of spillovers | | | | | | | |
|---|---|---|---|---|---|---|---|
| Sector group | O | | | TOTO | | | $\frac{\Sigma \text{TOTO}}{\Sigma \text{O}}$ (%) |
| | Min | Median | Max | Min | Median | Max | |
| Ke | 0.64 | 5.19 | 18.44 | 0.53 | 3.96 | 13.34 | 73.20 |
| Cg | 0.30 | 3.61 | 12.25 | 0.21 | 2.14 | 10.13 | 71.84 |
| Kg | 1.02 | 8.81 | 19.93 | 0.77 | 6.39 | 15.55 | 76.21 |
| Us | 0.56 | 6.15 | 15.69 | 0.41 | 4.46 | 11.16 | 75.59 |
| Panel B: sector groups as receivers of spillovers | | | | | | | |
| Sector group | I | | | TIFO | | | $\frac{\Sigma \text{TIFO}}{\Sigma \text{I}}$ (%) |
| | Min | Median | Max | Min | Median | Max | |
| Ke | 1.72 | 6.74 | 13.64 | 1.40 | 4.71 | 9.62 | 74.40 |
| Cg | 0.70 | 5.58 | 14.24 | 0.70 | 4.65 | 10.74 | 81.61 |
| Kg | 2.09 | 5.58 | 13.86 | 1.03 | 3.38 | 9.08 | 65.01 |
| Us | 2.75 | 5.48 | 16.97 | 1.65 | 4.37 | 11.94 | 74.33 |
| Panel C: the net spillovers and centrality indicators | | | | | | | |
| Sector group | Ri | | | WBC | | | |
| | Min | Median | Max | Min | Median | Max | |
| Ke | -0.86 | -0.13 | 0.52 | 0 | 18 | 120 | |
| Cg | -0.88 | -0.31 | 0.30 | 0 | 8 | 133 | |
| Kg | -0.46 | 0.20 | 0.68 | 0 | 33 | 235 | |
| Us | -0.79 | -0.06 | 0.64 | 0 | 27 | 208 | |

Table 5 reports the inter-group spillovers of the network in period 2. We can firstly find that both the total spillover intensity and the number of spillover paths in period 2 are significantly higher than those in period 1. It means that the outbreak of COVID-19 intensified the overall spillovers in China's stock market. The gross and net outward spillovers from Kg to Cg are 61.37 and 36.32, which account for higher proportion in total spillovers than those in period 1. Specifically, the proportion of the gross spillover from Kg to Cg increases from 9.1% to 9.5% of the total spillovers of the network. The proportion of the net spillover increases even more rapidly from 3.0% to 5.6%. Besides, the group Ke and Kg, as an integral whole, still has net spillover effect to the group Cg.

In period 2, the outward spillovers from Kg to other groups account for 24.0% of the total spillovers of the network. The group Cg receives 24.3% of the total spillovers. In addition, all other three groups have net outward spillovers to the group Cg, including the group Ke. Therefore, the Kg and Cg can be viewed as the major contributor and the receiver of the spillovers in period 2.

**Table5: The cross sector group analysis of volatility spillovers in period 2 (at the beginning of the CoVID-19 outbreak)**

| From\to | Intensity of spillovers | | | | | No. of spillover paths | | | | |
|---|---|---|---|---|---|---|---|---|---|---|
| | Ke | Cg | Kg | Us | Total | Ke | Cg | Kg | Us | Total |
| Ke | 42.48 | 46.01 | 31.08 | 36.65 | 646.31 | 277 | 348 | 230 | 234 | 4537 |
| Cg | 36.04 | 35.98 | 25.05 | 28.79 | | 267 | 328 | 261 | 252 | |

| | | | | | | | | | |
|---|---|---|---|---|---|---|---|---|---|
| Kg | 47.08 | 61.37 | 48.84 | 46.91 | | 301 | 403 | 273 | 275 |
| Us | 38.09 | 49.51 | 33.26 | 39.17 | | 214 | 313 | 222 | 245 |

**4.3 The group-specific spillover network in the second month after the CoVID-19 outbreak**

Table 6 reports the summary of the nodes in period 3 by group. According to Panel A, the sector group with the highest outward spillover effect in period 3 is the group Us, rather than the Kg in period 1 and period 2. The median of $O_i$ and that of $TOTO_i$ of group Us are 5.37 and 4.43. The group Us also has the highest $\sum TOTO/\sum O$ value, which is 76.21%.

According to Panel B, the median of $I_i$ of group Us is the highest, which is 5.76. However, the $\sum TIFO/\sum I$ value of group Us is not relatively high. It reveals that in period 3, the intra-group spillovers between the nodes in the group Us increase significantly. The group Ke still has the highest median of $TIFO_i$ (4.18). Moreover, the $\sum TIFO/\sum I$ value of group Ke is 78.41%.

According to Panel C, the group Kg has the highest median of $ri_i$ (0.05), which is still the only positive value. The lowest one belongs to the group Cg, which is -0.18. In period 3, the gap of $ri_i$ between different sector groups is closer than that in period 2. The group Us, instead of group Kg, becomes the group with the highest median of $WBC_i$, the value of which is 28. Besides, the median of $WBC_i$ of group Kg is 27, which is only slightly lower than that of the group Us.

**Table 6: The summary for indicators of the nodes by sector groups in period 3 (after the CoVID-19 outbreak preliminarily contained)**

| Panel A: sector groups as givers of spillovers | | | | | | | |
|---|---|---|---|---|---|---|---|
| Sector group | O | | | TOTO | | | $\frac{\sum TOTO}{\sum O}$ (%) |
| | Min | Median | Max | Min | Median | Max | |
| Ke | 0.35 | 4.70 | 9.87 | 0.35 | 3.76 | 8.53 | 75.10 |
| Cg | 0.23 | 3.95 | 25.20 | 0.17 | 2.52 | 18.68 | 69.67 |
| Kg | 1.54 | 5.19 | 15.51 | 1.29 | 4.22 | 11.76 | 76.93 |
| Us | 1.28 | 5.37 | 15.82 | 0.81 | 4.33 | 10.77 | 78.08 |
| Panel B: sector groups as receivers of spillovers | | | | | | | |
| Sector group | I | | | TIFO | | | $\frac{\sum TIFO}{\sum I}$ (%) |
| | Min | Median | Max | Min | Median | Max | |
| Ke | 1.41 | 5.42 | 12.05 | 1.30 | 4.18 | 9.06 | 78.41 |
| Cg | 0.84 | 5.46 | 11.61 | 0.56 | 3.82 | 8.60 | 72.99 |
| Kg | 2.05 | 5.58 | 11.50 | 1.22 | 4.10 | 7.26 | 75.07 |
| Us | 2.55 | 5.76 | 14.09 | 1.38 | 3.96 | 9.54 | 74.18 |
| Panel C: the net spillovers and centrality indicators | | | | | | | |
| Sector group | Ri | | | WBC | | | |
| | Min | Median | Max | Min | Median | Max | |
| Ke | -0.78 | -0.12 | 0.42 | 0 | 25 | 130 | |
| Cg | -0.91 | -0.18 | 0.60 | 0 | 9 | 242 | |
| Kg | -0.52 | 0.05 | 0.39 | 0 | 27 | 163 | |
| Us | -0.50 | -0.01 | 0.69 | 0 | 28 | 177 | |

Table 7 reports the inter-group spillovers of the network in period 3. Both the total spillover intensity and the number of spillover paths in period 3 are less than those in period 2. The most significant difference between the inter-group spillovers in period 3 and those in period 1 and period 2 is that net outward spillover effect from the group Us to the group Kg occurs. The group Us therefore becomes the group that has net outward spillovers to all other three groups. Besides, the group Ke and Kg, as an integral whole, still has net spillover effect to the group Cg.

In period 3, the outward spillovers from the group Us account for 20.4% in the total spillovers of the network. The inward spillovers to the group Cg account for 21.9% of the total spillovers. Therefore, the Us and Cg can be viewed as the major contributor and the receiver of the spillovers in period 3.

**Table 7: The cross sector group analysis of volatility spillovers in period 3 (after the CoVID-19 outbreak preliminarily contained)**

| From\to | Intensity of spillovers | | | | | No. of spillover paths | | | | |
|---|---|---|---|---|---|---|---|---|---|---|
| | Ke | Cg | Kg | Us | Total | Ke | Cg | Kg | Us | Total |
| Ke | 30.36 | 34.33 | 28.36 | 27.42 | 566.29 | 209 | 283 | 220 | 229 | 4254 |
| Cg | 39.40 | 47.55 | 35.03 | 32.77 | | 278 | 373 | 261 | 283 | |
| Kg | 35.40 | 41.61 | 33.25 | 32.19 | | 250 | 326 | 232 | 253 | |
| Us | 32.77 | 48.30 | 34.61 | 32.95 | | 211 | 305 | 206 | 220 | |

In conclusion, there are both stable patterns and significant changes in the demand-oriented industry-specific volatility spillover networks of China's stock market during the study period. Firstly, viewing the sector group Kg and Ke as an integral whole, they always have significant net spillovers to the group Cg. Such spillovers are relatively stronger in period 2 than in other two periods. Secondly, the group Kg always has the net outward spillover to the group Ke. More important, the importance of the group Us increasingly rises, and finally becomes the major contributor of the spillovers in period 3. Figure 2 depicts the simplified spillover paths of networks in different periods by sector groups.

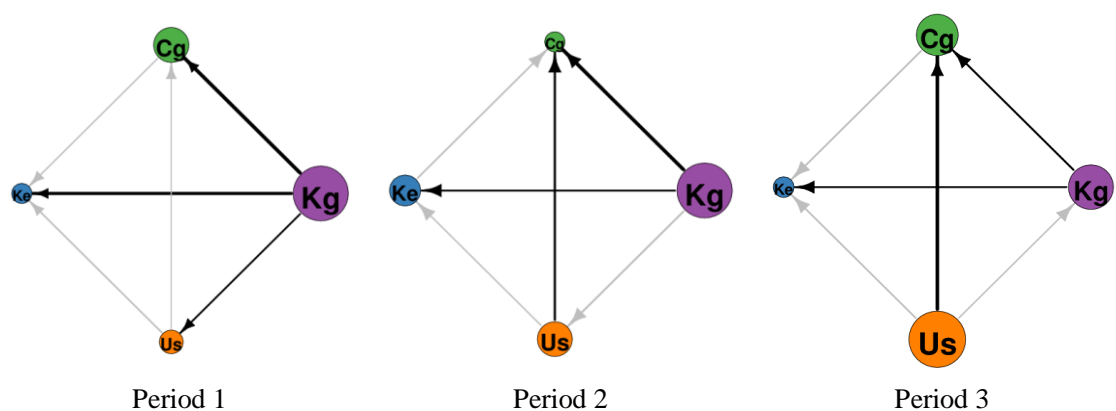

Period 1          Period 2          Period 3

Note: the size of each node represents the total spillover effects from this sector group to others. The width and direction of each arrow represent the strength and direction of net spillover effect between the relevant pair of sector groups, respectively. The black arrows in each subfigure represent the major paths in each period obtained through maximum spawning tree method.

**Figure 2: The demand-oriented industry-specific volatility spillover networks in different periods (net effect by sector groups)**

According to figure 2, the spillover paths from Kg to Cg and Ke are stable in all periods. The breakout of COVID-19 leads to an increasingly rise of importance of the spillover paths from the group Us to other ones. Especially, the path from Us to Cg becomes one of the major paths of the spillover networks of China's stock market after the breakout of COVID-19.

The findings of this section have inspiring economic implications. Firstly, some literatures like [56, 57] proved that fluctuation in investment demand caused by exogenous is the main cause of the fluctuation in China's economic demand. Our numerical findings further show that the structural change of volatility spillover networks of China's stock market can reflect the critical role that investment demand plays in the fluctuation in China's economic demand after the outbreak of COVID-19. On the one hand, the group Kg and Ke, as an integral whole, are the stable spillover contributors to the group Cg in the networks. On the other hand, the outward spillover effect from the groups Kg and Ke to the group Cg rapidly rose at the beginning of the COVID-19 outbreak.  After the outbreak was preliminarily contained, these spillovers significantly fell. One of the main damage caused by COVID-19 is the nationwide closure and idling of plants in all trades, which undisputedly has huge impact on the investment demand in China. The highly increased uncertainty of investment demand leads to the fluctuation in stock prices of securities in the industry sectors, which supply goods or services meeting the investment demand. Therefore, what is showed concerning the change in spillovers from the sector groups Ke and Kg to the group Cg provides the empirical evidence, from the perspective of financial market, for the economics theories purposed by [56, 57]. [56][57]

Secondly, the increasingly rising importance of the sector group Us in the spillover networks reveals the occurrence of supply restriction on service industry caused by the implementing of NPIs. For containing COVID-19, Chinese government implemented immediate NPIs nationwide. The majority part of the service places are forced to a termless shutting down. A large percentage of transportation services in China have to idle, despite for the huge freight and passenger traffic demands. The uncertainty of COVID-19 therefore transforms into the uncertainty of the operation environment of the companies of services sectors and then their asset prices. According to [58, 59], the service supply restriction will lead to the imbalance between supply and demand and retard economic growth. As the result, the companies of service sectors contribute more volatility spillovers to those of other sectors. When the overseas market demand is strong, a country is still able to achieve a high economic growth under the condition of service supply restriction. However, once the overseas market demand becomes insufficient, the service supply restriction will have serious damage to the economy. As introduced in section 2.1, the COVID-19 began to spread outside China in period 3. As a global pandemic, the COVID-19 will surely leads to the insufficiency of overseas demand of Chinese products. As the result, the importance of the group Us in the networks in period 3 is even higher than in period 2. In addition, according to [61], to compare with the period in which developed countries were at the similar stage of development as current China, there is severe supply restriction on most of service industries nowadays in China. Service supply restriction is an overall problem rather than a structural one in Chinese economy. The outbreak of COVID-19 is only an exogenous shock that intensified the problem. Therefore, we believe that our finding is still representative although not all service industry sectors are classified as members in the group Us according to the SWS standard. [58][59][60][61]

### 5. The further discussions

We further discuss the demand-oriented industry-specific volatility spillover networks of China's stock market in three aspects. First, we calculate the EMD of the distributions of the spillover intensity both inter- and intra- sector groups in different periods. Second, defining the major spillover paths as those with the top 20% highest intensity among a set of paths, we discuss the major spillover paths between different sector groups and their changes in different periods. Third, from various perspectives, we select the systemically important nodes of the networks in different periods.

**5.1 The EMDs between spillover intensity distributions in different periods**

Figure 3 depicts the intensity distributions of spillovers between sector groups in three periods. We can intuitively find that all of the intensity distributions of spillovers, both inter- and intra-sector groups, are right-skewed. Only few spillover paths have high intensity. In most of the subfigures, the intensity distribution of spillovers in period 1 is similar to that in period 3. The intensity distribution in period 2, in contrast, is significantly different from the counterparts in period 1 and of period 3. It reveals that in period 2, the industry-specific volatility spillover network of China's stock market has significantly structural changes. As an exception, the intensity distribution of spillover paths intra-group Us and of spillover paths between the group Us and Cg in period 2 are more similar to the distributions in period 3, rather than to the distributions in period 1. The exception is also consistent with the findings in section 4 and proves that the outbreak of COVID-19 has a more long-lasting impact on service sectors than on other ones.

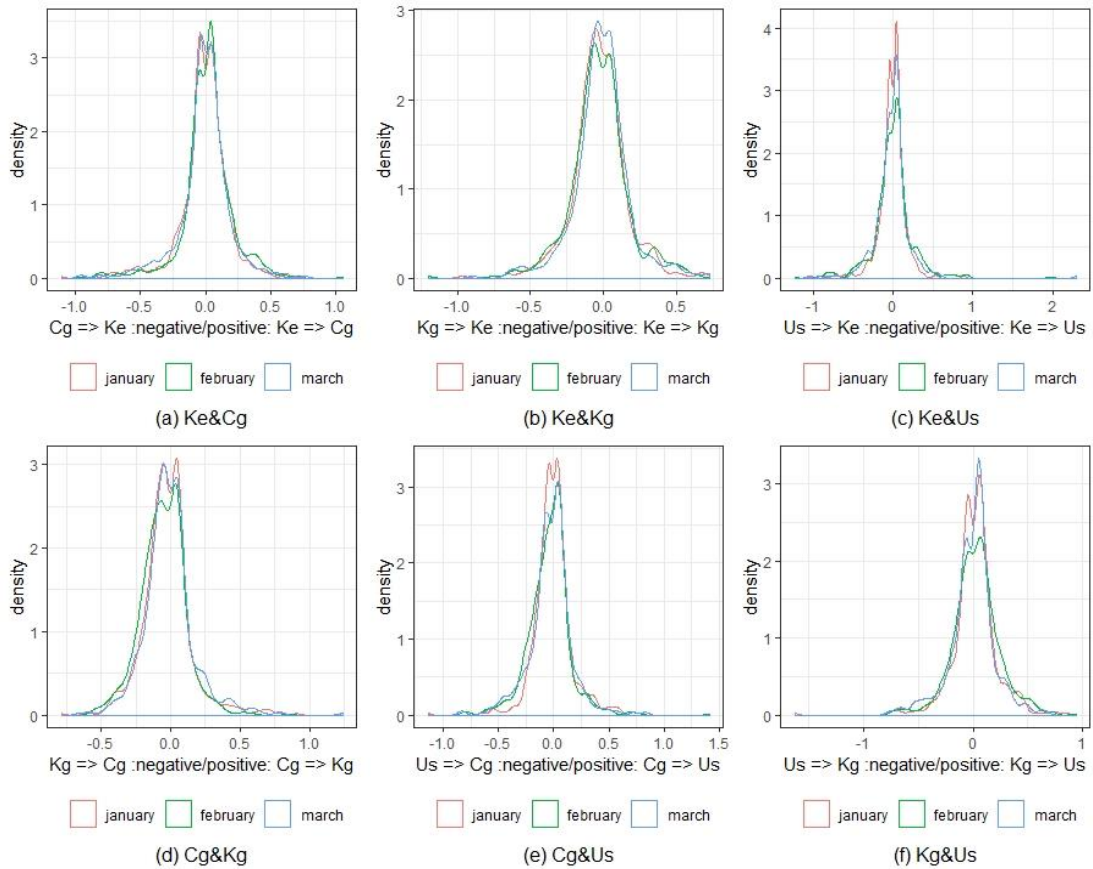

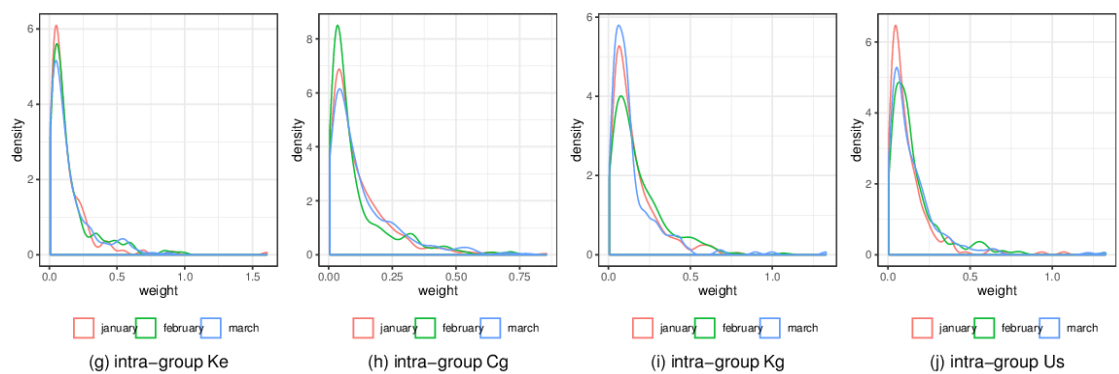

Note: From subfigure (a) to subfigure (f), to distinguish between the spillover paths in one direction to another, we do further processing with the data following the rule called "sector group B=>sector group A: negative/positive: sector group A=>sector group B". Take subfigure (a) for example, if a spillover path is from one of industries belonging to Ke group (which is viewed as group A) to one of those belonging to Cg group (which is viewed as group B), we take the original value of its intensity. Otherwise, we take the opposite number of its intensity.

**Figure 3 :The empirical probability density function of modified intensity of spillovers in different periods**

Table8 shows the EMDs between spillover intensity distributions in different periods. Between period 1 and period 2, most of high EMDs are connected with the distribution changes in spillovers between the group Us and other ones. Specifically, the EMD of the change in the intensity distribution of spillovers from Ke to Us is 6.34%, those from Us to Ke and to Cg are 5.11% and 4.38%, respectively. Between period 2 and period 3, most of high EMDs are connected with the distribution changes in spillovers between the group Kg and other ones. Specifically, being measured by the EMD, the intensity distribution of spillovers intra the group Kg changes in 4.43%. The EMD of the change in the intensity distribution of spillovers from Kg to Us and those from Cg to Kg are 4.31% and 3.71%, respectively.

**Table 8: EMDs between the intensity distributions of spillovers in different periods (%)**

| From\to | Period 1 vs. Period 2 | | | | Period 2 vs. Period 3 | | | |
|---|---|---|---|---|---|---|---|---|
| | Ke | Cg | Kg | Us | Ke | Cg | Kg | Us |
| Ke | 2.25 | 1.24 | 1.21 | 6.34 | 0.95 | 1.17 | 0.92 | 4.05 |
| Cg | 1.09 | 1.39 | 2.43 | 1.37 | 1.56 | 1.70 | 3.71 | 1.01 |
| Kg | 1.47 | 2.08 | 2.73 | 3.00 | 2.05 | 2.41 | 4.43 | 4.31 |
| Us | 5.11 | 4.38 | 2.37 | 3.70 | 2.53 | 1.39 | 1.90 | 1.40 |

The analysis in this section is a meaningful supplement for the analysis based on the sector influence indicator in section 4. From period1 to period 2, the group Us is the sector group of which the spillover effect strength distribution has the most significant change. It reveals that at the beginning of the outbreak of COVID-19, the nationwide implement of NPIs is reflected immediately in the distributional characteristics of the spillover networks of China's stock market. The significant distributional change in the spillover strength concerning the group Kg from period 2 to period 3, also shows that the risk for the investment demand destruction has been controlled to some extent. This is mainly owing to the successful contained pandemic of COVID-19 and the resumption of work that is strongly supported by both central and local governments of China. It means that the influence of the pandemic on the investment demand fell rapidly after the pandemic was contained in China, while the influence on the service sectors were rather more long-lasting.

**5.2 The major spillover paths**

Considering the right-skewed intensity distributions of spillovers showed in section 5.1, the spillover paths with high intensity are the minority but deserves further discussions. We define spillover paths with the top 20% highest intensity from sector group A to sector group B as the major spillover paths between these two groups. Figure 3 depicts the major spillover paths between different sector groups in different periods.

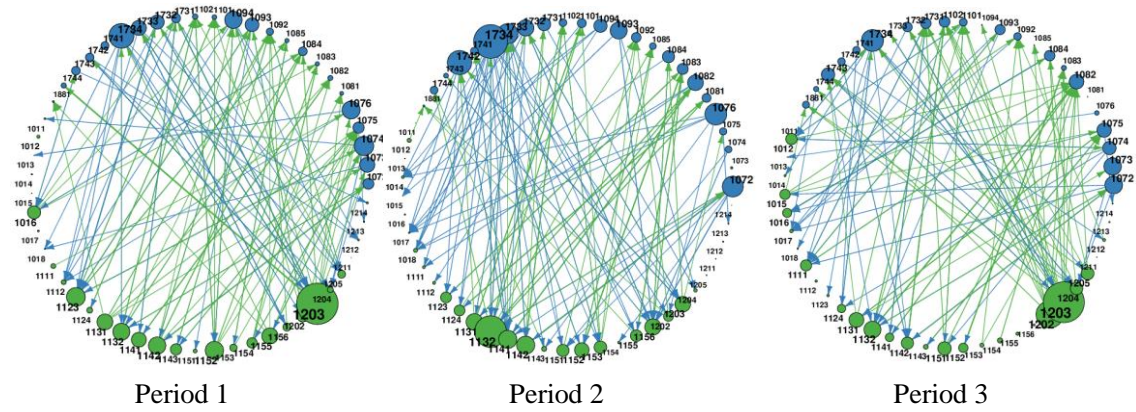

Period 1        Period 2        Period 3

(a) equipment manufacture v s. consumption goods

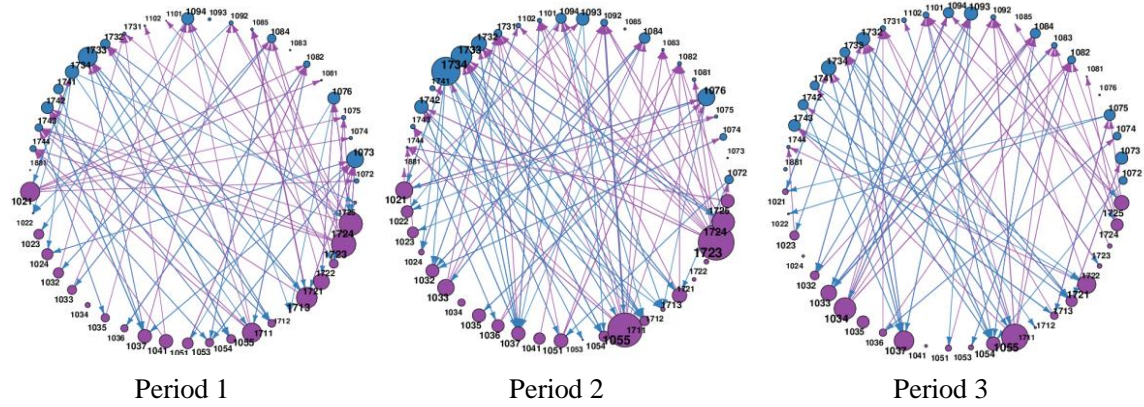

Period 1        Period 2        Period 3

(b) equipment manufacture v s. capital goods

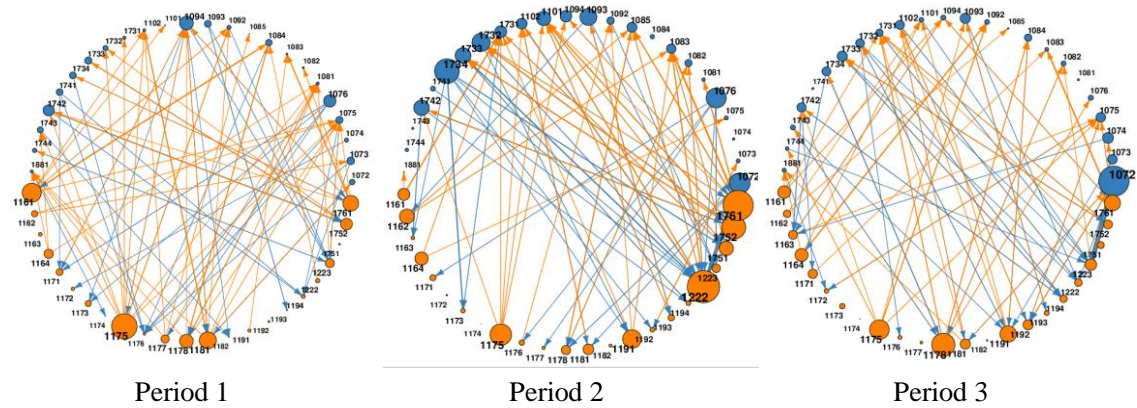

Period 1        Period 2        Period 3

(c) equipment manufacture v s. unclassified services

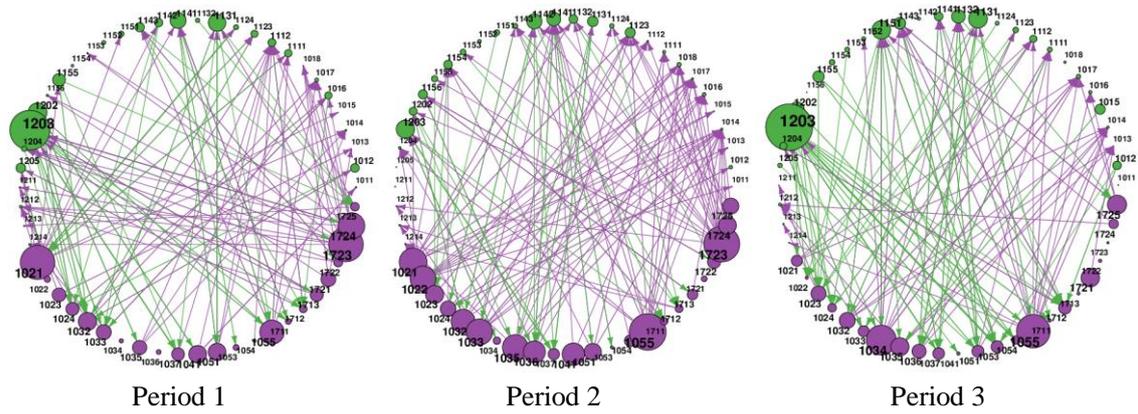

Period 1      Period 2      Period 3

(d) consumption goods v s. capital goods

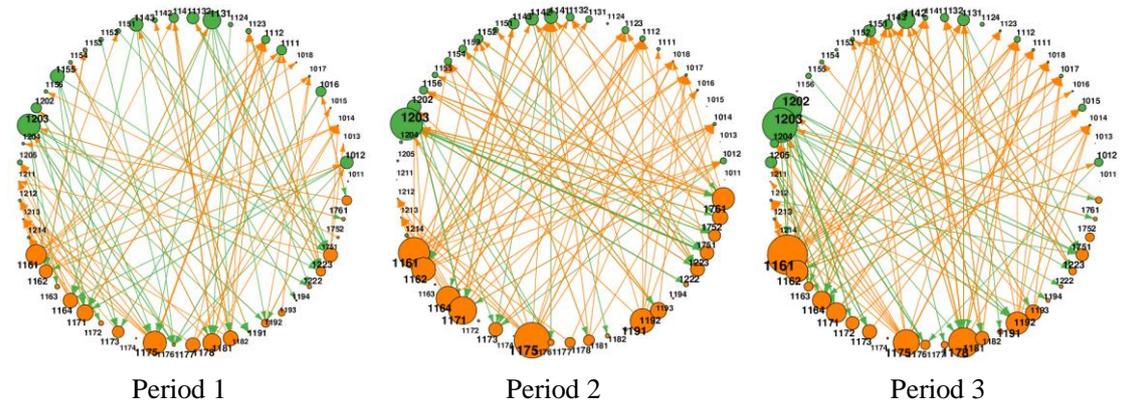

Period 1      Period 2      Period 3

(e) consumption goods v s. unclassified services

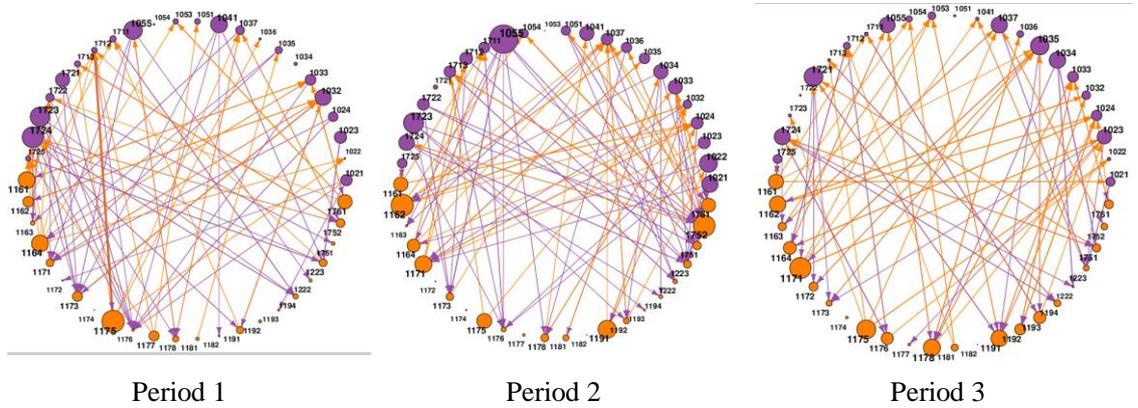

Period 1      Period 2      Period 3

(f) capital goods v s. unclassified services

Note: In all subfigures, the blue, green, purple and oranges nodes represent nodes of industry indices belonging to the sector group Ke, Cg, Kg and Us, respectively. The industry securities index name corresponding with each of the 4-digit codes in the figure can be found in Table 1.

**Figure 3 The major spillover paths inter-sector-group in different periods**

According to figure 3, the numbers of major spillover paths from the group Kg and Ke to other groups increase from period 1 to period 2, while decrease from period 2 to period 3. The number of major spillover paths from the group Cg, in contrast, decreases from period 1 to period 2, while increases from period 2 to period 3. The number of major spillover paths from the group Us increases from period 1 to period 2. However, this number does not change significantly from period 2 to period 3.

Specifically, according to subfigure (a) (b) and (c), in the sector group Ke, the electrical equipment sector (1731, 1732, 1733 and 1734) is one of the critical contributors of the major spillovers to other sector groups in the whole study period. According to subfigure (a) (d) and (e), the major contributors of spillovers from sector group Cg vary depending on the periods. In period 1, the commercial trade sector (1202, 1203, 1204 and 1205), the apparel and textiles sector (1131, 1132) and the light manufacturing sector (1141, 1142 and 1143) have relatively large numbers of major spillover paths to other sector groups. In period 2 and period 3, the commercial trade sector still plays the main contributor of major spillover paths, while the spillovers from the other two sectors become far less intensive than in period 1. Besides, the biochemical and pharmaceuticals sector (indices from 1151 to 1156) becomes a new critical spillover contributor. According to subfigure (b) (d) and (f), in period 1 the construction and decorations sector (1711, 1712 and 1713) and building materials (1721, 1722, 1723, 1724 and 1725) contribute most of major spillover paths to other sector groups. However, various sectors of mining and raw material processing become more critical spillover contributors in period 2 and period 3. According to subfigure (c) (e) and (f), in period 1, the number of major spillover paths from utilities sector (indices from 1161 to 1164) and transporting sector (indices from 1171 to 1178) account for the largest proportion of all major spillover paths from the group Us to other sector groups. The numbers of major spillover paths from the information services sector (1222, 1223) and the media sector (1751, 1752 and 1761) increase in period 2. The number of major spillover paths from the finance sector increases in period 3.

Figure 4 depicts the major spillover paths intra-sector-group in different periods.

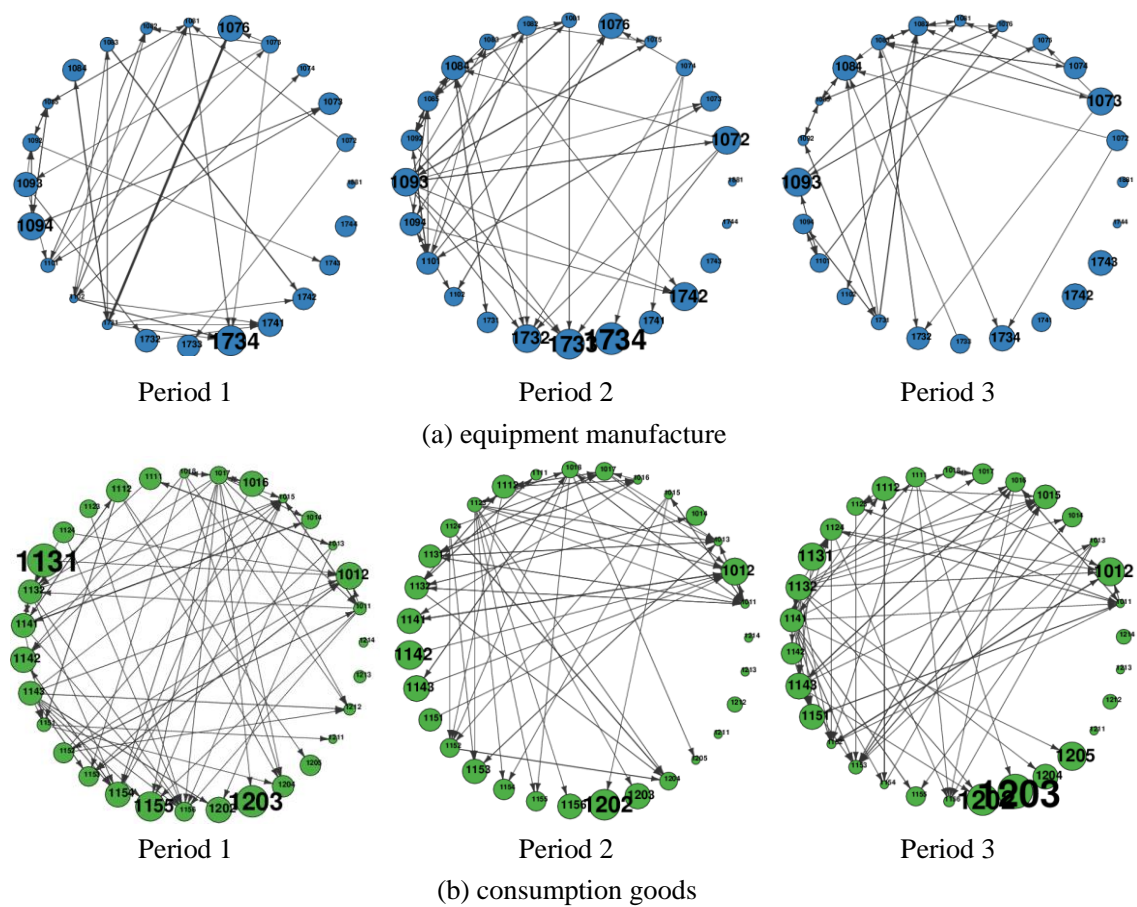

Period 1　　　　　　　　　　Period 2　　　　　　　　　　Period 3
(a) equipment manufacture

Period 1　　　　　　　　　　Period 2　　　　　　　　　　Period 3
(b) consumption goods

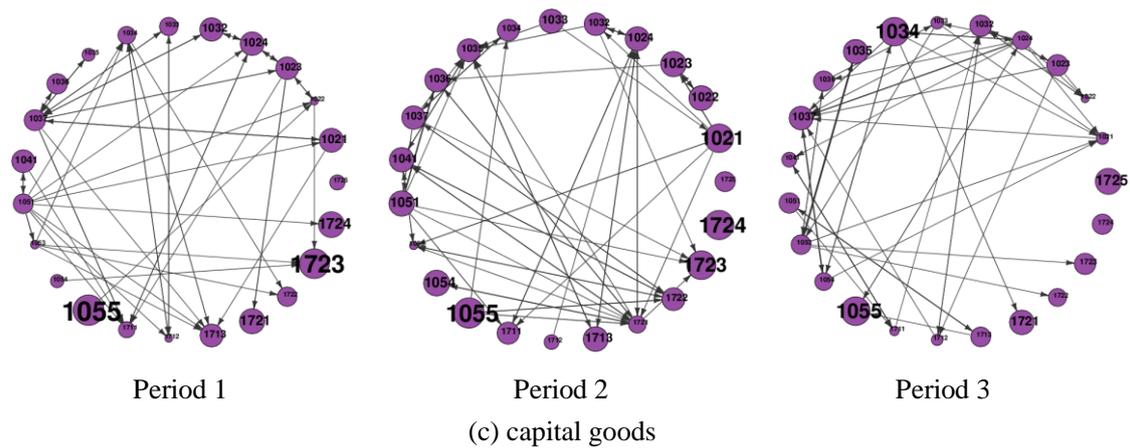

| Period 1 | Period 2 | Period 3 |

(c) capital goods

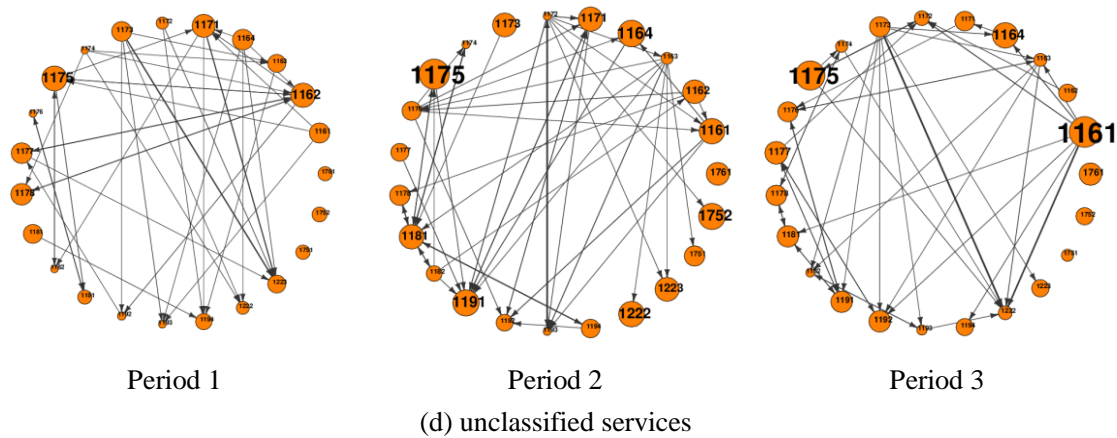

| Period 1 | Period 2 | Period 3 |

(d) unclassified services

Note: in all subfigures, the blue, green, purple and oranges nodes represent vertices of industry indices belonging to the sector group Ke, Cg, Kg and Us, respectively. The industry securities index name corresponding with each of the 4-digit codes in the figure can be found in appendix.

**Figure 4 The major spillover paths intra-sector-group in different periods**

According to figure 4, the numbers of major spillover paths intra the group Kg and intra the group Ke increase from period 1 to period 2, while decrease from period 2 to period 3. The number of major spillover paths intra the group Cg, in contrast, decreases from period 1 to period 2, while increases from period 2 to period 3. The number of major spillover paths intra the group Us does not change significantly during the whole study period.

Specifically, according to subfigure (a), there are large number of major spillover paths between the transportation equipment sector (1092, 1093, 1094 and 1881), the machinery sector (from 1072 to 1076) and the electronic components sector (from 1081 to 1085). There a great number of major spillover paths intra the group Ke received by the electrical equipment sector during the whole study period. This number is especially great in period 2. According to subfigure (b), in the sector group Cg, the apparel and textiles sector and the light manufacturing sector contribute a great number of major spillover paths to the commercial trade sector, the agricultural sector (from 1011 to 1018) and the biochemical and pharmaceuticals sector in period 1. After period 2, the major spillovers received by agricultural sector significantly increases, while the numbers of those received by the commercial trade sector and the biochemical and pharmaceuticals sector decrease. According to subfigure (c), in the sector group Kg, the mining sector and the raw material processing sectors have a number of major spillover paths to the construction sectors. In addition, the number of major spillover paths from the mining sector and the raw material processing sectors to the construction sectors increases in to period 2, while decreases in period 3. According to subfigure (d), in sector group Us, the number of major spillover paths from the transportation service sector and the utilities sector to other sectors are great, and it is even greater in period 2 than in period 1 or period 3. Moreover, the finance sector has more major spillover paths to other sectors in group Us after period 2.

**5.3 The systematically important nodes in spillover network**

The structure of the volatility spillover networks of China's stock market is asymmetric in the study period.

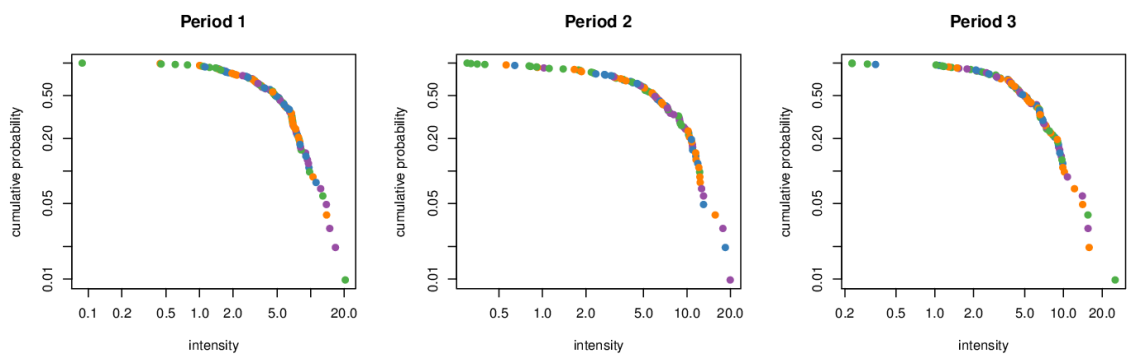

(a) the outward connectivity of the nodes

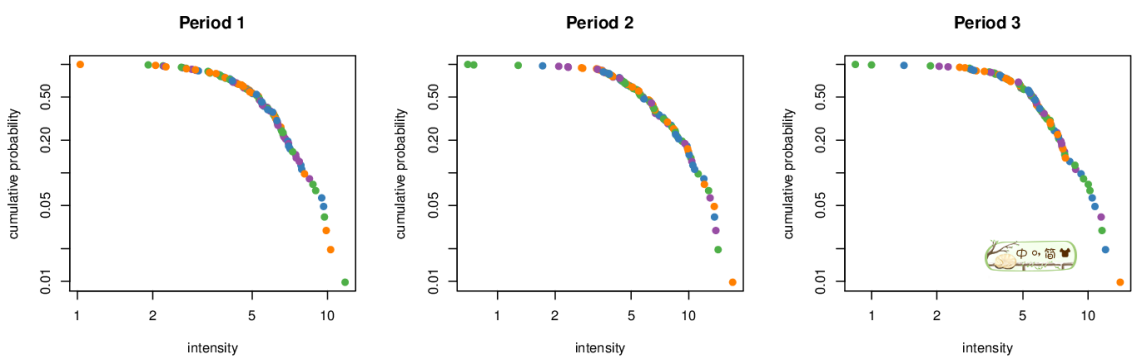

(b) the inward connectivity of the nodes

Note: The blue, green, purple and oranges nodes represent vertices of industry indices belonging to the sector group Ke, Cg, Kg and Us, respectively. The conglomerates industry securities index (801231) is represented by a grey node.

**Figure 5 The distributions of the nodes' connectivity intensity in different periods**

According to figure 5, although the exact distribution of the nodes' connectivity intensity cannot be examined, we can still find that the several nodes with the highest connectivity intensity play the dominated role in the spillover networks. Thus, we further select the systematical important nodes in the volatility spillover networks in different periods from various perspectives.

Firstly, according to Figure 6 (a), nodes in the commerce and trading sector (1202 and 1203), the construction sector (1721, 1723 and 1724), the utility sector (1161) and the transportation services sector (1175 and 1178) can be viewed as the main contributors of the spillovers in all periods. Comparing to the network in period 1, in the spillover network in period 2, there are more nodes in the group Ke and in the group Us that become the main contributors of spillover effects, while less nodes in the group Cg that keep playing the role of main contributors. Comparing to the network in period 2, in the spillover network in period 3, there are more nodes in the group Cg and in group Us that become the main contributors of spillovers, while less nodes in group Ke and in group Kg that keep playing the role of main contributors.

According to figure 6(b), a list of nodes in sector group Cg (1014, 1017, 1111, 1112, 1212, 1123, 1141, 1143 and 1156) and in the group Us (1171, 1176, 1123, 1223 and 1752) are the main receivers of spillovers in the networks during the whole study period. After period 2, more nodes in sector group Kg (1712, 1037 and 1053) and in the group Ke (1731, 1732, 1733, 1101 and 1084) become the main receivers. Based on figure 7, figure 6 further provides the visualization of the relative spillovers of the nodes according to their relative influence indicators.

According to figure 8, comparing to the network in period 1, in the network in period 2, there are more nodes in the group Kg and Ke that have the relatively higher betweenness centrality and k-shell level. However, in the network in period 3, there are more nodes in group Us that have the relatively higher betweenness centrality and k-shell level.

In conclusion, the analysis in section 5 further validates the main result of section 4. The spillovers from sector groups Kg and Ke, as an integral whole, rise in period2, while fall in period 3. The change in the spillovers from the group Cg is in contrast. For the opinion of the spillovers from sectors meeting the consumption demand to other sectors in China's stock market, our findings are consistent with literature [45] (in Chinese). Both the literature [45] and this paper are early reports of the industry-specific volatility spillover networks of china's stock market around the outbreak of COVID-19. Comparing to [45], however, we select the SWS standard, rather than the CSI standard, to identify the networks. The relationship between the sectors meeting the investment and consumption demand therefore can be further detected. as the result, we can illustrate how spillover networks of China's stock market reflect the relative rise and fall of the uncertainty of the investment demand and consumption demand in China during the spread of COVID-19. Literature [45] also purposed the proposition that the services industries suffering from the COVID-19 is a potential threat of the economy recovery in China. Besides, literature [42] found that more difficulties would be faced by industry sectors relying on transportation services in the economy recovery. Basing on the spillover networks of stock market, our findings provide evidence for these literatures.

**6. Conclusion**

To illustrate how spillover networks of stock market reflect the change in demand, we identified the GARCH-BEKK based demand-oriented industry-specific volatility spillover networks of China's stock market by dividing 102 SWS securities indices into four demand-oriented sector groups. We divided the study period into three different sub-periods. Period 1, 2 and 3 represent the period before the nationwide outbreak of COVID-19 in China, the period at the beginning of the outbreak and the one after the disease was preliminarily contained.

According to our empirical analysis, firstly, the sector groups Ke and Kg, as a whole, have stable net spillovers to the sector group Cg during the whole study period. Secondly, the net spillovers from the group Ke and Kg to the group Cg rose in period 2, while fell in period 3. Thirdly, as of period 2, the importance of the sector group Us became increasingly higher. The group Us finally played the main contributor of the spillover network of China's stock market in period 3. We did further discussions from various perspectives. All discussions did a joint validation for our main result.

We emphasize discussing the demand change in a country. Our findings also have meaningful inspirations about economy recovery in the context of containing the spread of COVID-19. The investment demand in China suffered more than the consumption demand did from the exogenous shock of the COVID-19 at the beginning of outbreaks of this disease. However, when the pandemic was contained, the risk in investment demand in China was also controlled to some extent. On the one hand, fluctuation in investment demand caused by exogenous shocks is the main reason of the uncertainty of economy in one country. It is therefore consistent with the economics theories that the sector groups meeting the investment demand dominate the spillover network of China's stock market at the beginning of the outbreak of COVID-19. On the other hand, because Chinese government successfully contained the pandemic and organized the work resumption with high efficiency, the risk of investment demand was controlled and eliminated relatively fast. The trend of risk eliminating is also reflected in the structural change of the spillover networks between period 2 and period 3.

The sector group US began to play critical role in the spillover network as of the nationwide implement of NPIs. The importance of the sector group Us rose even higher when the COVID-19 deteriorated out of China. It reveals that the supply restriction in services is still a long-lasting threat for the Chinese economic recovery in the next stage, especially under the condition that foreign demand is destroyed by COVID-19. We believe that the NPIs is necessary for all countries and regions suffering from the COVID-19. Thus, being aware of the overall influence of services sectors is critical for investors and policymakers all over the world.

Finally, for further research, more proper industry classification standards can be established for or introduced into the spillover network analysis on different research targets. Exogenous variables should be introduced into the industry-specific spillover networks of capital markets. Then we can do more quantitative analysis on the risk contagion and spillover effects between the exogenous shock and the securities of companies belonging to different industry sectors.

Appendix A

**Table A-1 The industry securities index names corresponding with their index codes**

| codes | Index names | codes | Index names |
|---|---|---|---|
| 801011 | Forestry | 801155 | Chinese medicine |
| 801012 | Agricultural Products | 801156 | Health Care Service |
| 801013 | Agricultural Conglomerates | 801161 | electric Utilities |
| 801014 | Feed Processing | 801162 | Environmental Facilities & Service |
| 801015 | Fishery | 801163 | Gas Utilities |
| 801016 | Farming | 801164 | Water Utilities |
| 801017 | Husbandry | 801171 | Marine Ports & Service |
| 801018 | Animal Health | 801172 | public transit |
| 801021 | Coal Mining | 801173 | Airlines |
| 801022 | other Mining | 801174 | Airport service |
| 801023 | Oil & Gas Drilling | 801175 | Highways |
| 801024 | Mining Equipment & Services | 801176 | Marine |
| 801032 | chemical fiber | 801177 | Railroads |
| 801033 | Chemical materials | 801178 | Trucking |
| 801034 | chemical products | 801181 | Real Estate Management & Development |
| 801035 | Petrochemical Industry | 801182 | Park Exploitation |
| 801036 | Plastic | 801191 | Diversified Financial Service |
| 801037 | Rubber | 801192 | Banks |
| 801041 | Steel | 801193 | Capital Markets |
| 801051 | Metal New materials | 801194 | Insurance |
| 801053 | Gold | 801202 | Trading |
| 801054 | Precious Metals & Minerals | 801203 | retailing |
| 801055 | Industrial Metal | 801204 | Specialty Retail |
| 801072 | General Industrial Machinery | 801205 | Commercial Property Service |
| 801073 | Instrument & Apparatus | 801211 | Catering |
| 801074 | Special Equipment | 801212 | Attractions |
| 801075 | Metal Products | 801213 | Hotel |
| 801076 | Transporting Facilities | 801214 | Leisure Conglomerates |
| 801081 | Semi-conductor | 801222 | Software |
| 801082 | Other Electronic Products | 801223 | IT Services |
| 801083 | Electronical Part & Component | 801231 | Conglomerates |
| 801084 | Optical & Opto-electronic Products | 801711 | Cement |
| 801085 | Electronical Manufacturing | 801712 | Glass Products |
| 801092 | Automobile Services | 801713 | Other Construction Materials |
| 801093 | Auto Parts & Equipment | 801721 | Homebuilding |
| 801094 | Automobile Manufacturers | 801722 | Decoration |
| 801101 | Computers & Peripherals | 801723 | Infrastructures |
| 801102 | Communications Equipment | 801724 | Specialty Engineering |
| 801111 | Household Appliances | 801725 | Landscape engineering |
| 801112 | Audiovisuals | 801731 | electrical machinery |
| 801123 | Beverage | 801732 | Electric Automation Equipment |
| 801124 | Food Products | 801733 | power supply equipment |
| 801131 | Textiles | 801734 | High-Low-voltage Switch Equipment |
| 801132 | Apparel | 801741 | Aerospace Equipment |
| 801141 | Packaging & Printing | 801742 | Aviation Equipment |
| 801142 | Household Products | 801743 | Defense Equipment |
| 801143 | Paper Products | 801744 | Shipbuilding |
| 801151 | Chemical pharmacy | 801751 | Advertising & Broadcasting |
| 801152 | Biotechnology | 801752 | Internet Media |
| 801153 | Health Care Equipment | 801761 | Culture Media |
| 801154 | Health Care Distributors | 801881 | Other Transporting Equipment |

Appendix B

**Table B-1 Summary of minute-per-minute returns of the SWS securities industry indices**

| Index | Mean(‰) | SD (%) | Skew | Kurt | JB test | AR1 | ADF |
|---|---|---|---|---|---|---|---|
| 801011 | -0.903 | 16.065 | 0.746 | 12.393 | 13395*** | -0.312*** | -15.291*** |
| 801012 | -0.681 | 5.323 | -0.205 | 6.578 | 1921*** | -0.167*** | -14.535*** |
| 801013 | -0.81 | 16.37 | 0.212 | 8.831 | 5061*** | -0.245*** | -16.544*** |
| 801014 | -1.978 | 11.613 | 1.438 | 39.187 | 195139*** | -0.027*** | -15.595*** |
| 801015 | -0.343 | 7.478 | -0.501 | 8.214 | 4175*** | -0.292*** | -14.733*** |
| 801016 | -0.169 | 7.627 | 7.071 | 241.454 | 8449653*** | 0.011*** | -15.491*** |
| 801017 | -2.167 | 12.145 | 1.729 | 44.891 | 261638*** | -0.048*** | -14.903*** |
| 801018 | -1.939 | 9.717 | -0.05 | 12.299 | 12805*** | -0.115*** | -15.709*** |
| 801021 | -0.169 | 5.56 | -0.977 | 37.747 | 179355*** | 0.003*** | -14.786*** |
| 801022 | 1.28 | 11.014 | -0.096 | 5.133 | 679*** | -0.344*** | -16.618*** |
| 801023 | 1.131 | 11.162 | 0.001 | 3.123 | 2 | -0.452*** | -17.112*** |
| 801024 | 1.362 | 7.237 | 0.04 | 8.987 | 5308*** | -0.142*** | -13.563*** |
| 801032 | 0.436 | 5.511 | 0.157 | 9.999 | 7268*** | -0.066*** | -14.094*** |
| 801033 | -0.763 | 5.942 | -0.425 | 10.867 | 9271*** | -0.136*** | -14.543*** |
| 801034 | -0.034 | 4.086 | -2.999 | 111.755 | 1756809*** | 0.155*** | -14.632*** |
| 801035 | 0.543 | 8.122 | -0.022 | 3.004 | 0 | -0.433*** | -15.593*** |
| 801036 | 0.146 | 6.713 | 3.799 | 114.012 | 1833495*** | -0.053*** | -15.153*** |
| 801037 | 0.301 | 5.605 | -0.742 | 15.213 | 22414*** | -0.11*** | -14.207*** |
| 801041 | 0.951 | 9.368 | 0.377 | 24.26 | 67013*** | -0.201*** | -14.351*** |
| 801051 | -0.608 | 5.895 | -1.513 | 29.834 | 107985*** | 0.054*** | -14.748*** |
| 801053 | 1.742 | 11.145 | -0.355 | 70.667 | 678131*** | -0.153*** | -15.589*** |
| 801054 | 2.997 | 7.955 | 2.589 | 40.927 | 216983*** | 0.062*** | -15.215*** |
| 801055 | -0.042 | 5.526 | -0.366 | 7.556 | 3154*** | -0.233*** | -13.098*** |
| 801072 | -0.799 | 4.243 | -1.937 | 76.396 | 799947*** | 0.084*** | -14.131*** |
| 801073 | 3.452 | 6.1 | 1.167 | 25.485 | 75674*** | 0.018*** | -15.159*** |
| 801074 | -1.047 | 4.677 | -2.233 | 58.968 | 466814*** | 0.06*** | -14.793*** |
| 801075 | 0.098 | 5.48 | -0.329 | 8.875 | 5175*** | -0.141*** | -13.245*** |
| 801076 | -0.068 | 7.743 | 0.124 | 6.562 | 1888*** | -0.332*** | -15.222*** |
| 801081 | 0.45 | 10.877 | -0.902 | 26.515 | 82365*** | 0.105*** | -13.42*** |
| 801082 | 0.194 | 7.866 | 0.627 | 11.823 | 11760*** | 0.077*** | -14.136*** |
| 801083 | -1.317 | 9.022 | -1.62 | 33.36 | 138049*** | 0.068*** | -13.503*** |
| 801084 | -0.144 | 8.405 | -1.465 | 30.54 | 113585*** | 0* | -14.966*** |
| 801085 | 0.205 | 9.623 | -0.513 | 24.526 | 68771*** | 0.081*** | -14.303*** |
| 801092 | 0.746 | 12.032 | -0.273 | 6.601 | 1964*** | -0.359*** | -16.821*** |
| 801093 | 0.464 | 5.12 | -1.222 | 75.057 | 769773*** | 0.053*** | -15.378*** |
| 801094 | -0.855 | 5.687 | -0.783 | 28.128 | 93863*** | -0.021*** | -16.051*** |
| 801101 | -1.415 | 7.657 | -2.453 | 55.807 | 416502*** | 0.1*** | -14.612*** |
| 801102 | -1.525 | 6.582 | -3.256 | 105.064 | 1548860*** | 0.165*** | -14.506*** |
| 801111 | -1.699 | 7.908 | -4.988 | 116.867 | 1934734*** | 0.082*** | -13.435*** |
| 801112 | 0.67 | 8.655 | -0.004 | 6.449 | 1762*** | -0.261*** | -13.704*** |
| 801123 | 0.02 | 5.659 | 0.076 | 14.317 | 18970*** | 0.101*** | -14.104*** |
| 801124 | -0.563 | 6.134 | -0.148 | 17.297 | 30282*** | -0.025*** | -14.013*** |
| 801131 | -0.056 | 4.318 | -0.34 | 19.892 | 42321*** | -0.108*** | -13.335*** |
| 801132 | -1.593 | 5.244 | -4.632 | 115.477 | 1886105*** | -0.049*** | -14.272*** |
| 801141 | -0.478 | 5.487 | -0.3 | 9.228 | 5797*** | -0.002*** | -14.648*** |
| 801142 | -0.532 | 4.744 | -1.367 | 45.758 | 271836*** | 0.025*** | -14.31*** |
| 801143 | -0.433 | 6.842 | -0.82 | 10.068 | 7796*** | -0.112*** | -15.213*** |
| 801151 | 0.74 | 5.486 | 0.174 | 45.784 | 271083*** | 0.119*** | -15.87*** |
| 801152 | -1.356 | 6.436 | -1.093 | 25.637 | 76590*** | 0.161*** | -12.885*** |
| 801153 | -1.334 | 6.454 | -1.359 | 26.376 | 82010*** | 0.16*** | -12.518*** |
| 801154 | -1.427 | 4.945 | 0.03 | 12.11 | 12290*** | -0.026*** | -13.718*** |
| 801155 | -1.345 | 4.754 | -2.261 | 53.329 | 378124*** | 0.16*** | -14.2*** |
| 801156 | -2.577 | 8.67 | -0.426 | 17.912 | 33035*** | 0.09*** | -12.896*** |
| 801161 | -0.411 | 4.091 | -0.905 | 24.588 | 69499*** | -0.161*** | -14.244*** |
| 801162 | -0.152 | 4.254 | -0.526 | 19.486 | 40411*** | 0.012*** | -12.533*** |
| 801163 | -0.31 | 5.695 | -0.632 | 23.393 | 61819*** | -0.038*** | -15.138*** |
| 801164 | -0.563 | 5.913 | -0.219 | 5.001 | 622*** | -0.338*** | -13.14*** |
| 801171 | -0.084 | 6.469 | -0.147 | 4.522 | 356*** | -0.354*** | -13.978*** |
| 801172 | 0.622 | 6.596 | -0.309 | 20.147 | 43595*** | -0.274*** | -14.805*** |
| 801173 | 0.619 | 9.279 | -0.044 | 5.41 | 861*** | -0.343*** | -14.24*** |
| 801174 | 0.729 | 7.313 | 0.034 | 11.978 | 11938*** | -0.076*** | -12.674*** |
| 801175 | 0.723 | 3.825 | -0.373 | 6.507 | 1904*** | -0.279*** | -13.566*** |
| 801176 | 1.587 | 8.194 | 0.751 | 12.524 | 13767*** | -0.169*** | -13.217*** |
| 801177 | 0.174 | 8.336 | 0.144 | 5.303 | 797*** | -0.403*** | -15.965*** |
| 801178 | -1.142 | 4.751 | -1.183 | 26.155 | 80226*** | -0.044*** | -14.08*** |
| 801181 | -0.869 | 4.372 | -1.865 | 61.322 | 505764*** | 0.075*** | -14.24*** |
| 801182 | -0.189 | 5.001 | -0.036 | 50.9 | 339765*** | -0.117*** | -13.432*** |
| 801191 | -0.07 | 6.451 | -0.567 | 23.318 | 61325*** | -0.087*** | -14.984*** |
| 801192 | -0.726 | 4.571 | 0.141 | 15.213 | 22100*** | -0.124*** | -15.312*** |
| 801193 | -0.799 | 6.114 | 1.626 | 58.392 | 455934*** | 0.08*** | -14.999*** |
| 801194 | -1.106 | 5.455 | 0.029 | 20.385 | 44756*** | 0.027*** | -15.165*** |
| 801202 | 0.073 | 5.736 | -0.133 | 14.933 | 21097*** | -0.234*** | -14.441*** |
| 801203 | 0.529 | 4.123 | -0.237 | 21.943 | 53169*** | -0.104*** | -13.023*** |
| 801204 | -0.132 | 6.666 | 0.083 | 10.175 | 7627*** | -0.214*** | -15.7*** |
| 801205 | 0.129 | 7.235 | 0.473 | 15.382 | 22836*** | -0.29*** | -15.406*** |
| 801211 | -3.334 | 11.18 | 0.032 | 6.149 | 1469*** | -0.199*** | -15.77*** |
| 801212 | -0.138 | 8.416 | 0.541 | 16.227 | 26081*** | -0.216*** | -15.32*** |
| 801213 | 3.7 | 10.659 | 1.261 | 22.761 | 58767*** | -0.067*** | -13.902*** |
| 801214 | -1.14 | 9.226 | -0.885 | 34.51 | 147491*** | -0.023*** | -14.829*** |
| 801222 | -0.788 | 6.87 | -2.365 | 79.895 | 878908*** | 0.165*** | -14.683*** |
| 801223 | 0.796 | 11.852 | 0.029 | 6.514 | 1829*** | -0.352*** | -15.129*** |
| 801231 | -0.16 | 5.522 | -0.291 | 34 | 142354*** | -0.1*** | -14.273*** |
| 801711 | 1.899 | 8.488 | 3.049 | 77.688 | 831555*** | 0.12*** | -15.411*** |
| 801712 | 1.441 | 8.734 | 0.216 | 7.709 | 3311*** | -0.252*** | -15.216*** |
| 801713 | 0.895 | 5.733 | 0.907 | 22.886 | 59051*** | 0.016*** | -15.016*** |
| 801721 | 0.191 | 9.769 | 0.134 | 4.434 | 315*** | -0.382*** | -15.273*** |
| 801722 | 0.785 | 5.303 | -1.06 | 22.936 | 59520*** | -0.116*** | -16.152*** |
| 801723 | 0.678 | 5.343 | 0.459 | 15.852 | 24586*** | -0.093*** | -12.759*** |
| 801724 | 0.836 | 6.232 | -0.05 | 6.752 | 2086*** | -0.3*** | -13.615*** |
| 801725 | -1.986 | 6.082 | -0.631 | 21.37 | 50208*** | -0.222*** | -14.937*** |
| 801731 | 1.57 | 8.162 | -0.619 | 18.209 | 34482*** | -0.108*** | -15.183*** |
| 801732 | 0.149 | 6.248 | -0.255 | 12.16 | 12463*** | -0.045*** | -14.955*** |
| 801733 | 2.507 | 6.899 | 6.863 | 213.396 | 6583022*** | 0.032*** | -14.506*** |
| 801734 | -0.391 | 4.735 | -1.798 | 57.386 | 439925*** | -0.015*** | -14.019*** |
| 801741 | 1.939 | 5.806 | 2.174 | 50.854 | 341915*** | -0.065*** | -14.438*** |
| 801742 | 1.342 | 4.438 | -0.222 | 13.912 | 17663*** | 0.096*** | -13.931*** |
| 801743 | 1.086 | 5.518 | 0.065 | 16.33 | 26314*** | -0.145*** | -15.402*** |
| 801744 | -0.54 | 9.756 | 0.483 | 98.549 | 1352077*** | -0.143*** | -16.34*** |
| 801751 | -0.737 | 8.929 | -0.352 | 7.031 | 2480*** | -0.176*** | -14.61*** |
| 801752 | 1.823 | 7.355 | -1.207 | 45.102 | 263352*** | 0.132*** | -13.324*** |
| 801761 | -1.072 | 4.977 | -1.975 | 36.741 | 170898*** | 0.034*** | -13.108*** |
| 801881 | 0.67 | 7.784 | 0.006 | 7.847 | 3479*** | -0.189*** | -15.836*** |

Note: Descriptive statistics of the 102 industry indices' minute-per-minute returns includes means, standard deviations (SD), skewness (Skew) and kurtosis (Kurt). The results of Jarque-Bera normality test (JB test), the first order autocorrelation (AR1) and the augmented Dickey–Fuller test (ADF) are also reported. *, **, and *** denote the rejection of the null hypothesis at the 10%, 5%, and 1% level, respectively. The null hypothesis for the AR1, the JB tests, and the ADF tests is that the first-order autocorrelation is zero. In another word, the series is normally distributed, and has a unit root. The number of observations counts 12084.

Appendix C  The figure 6, figure 7, and figure 8 in the text

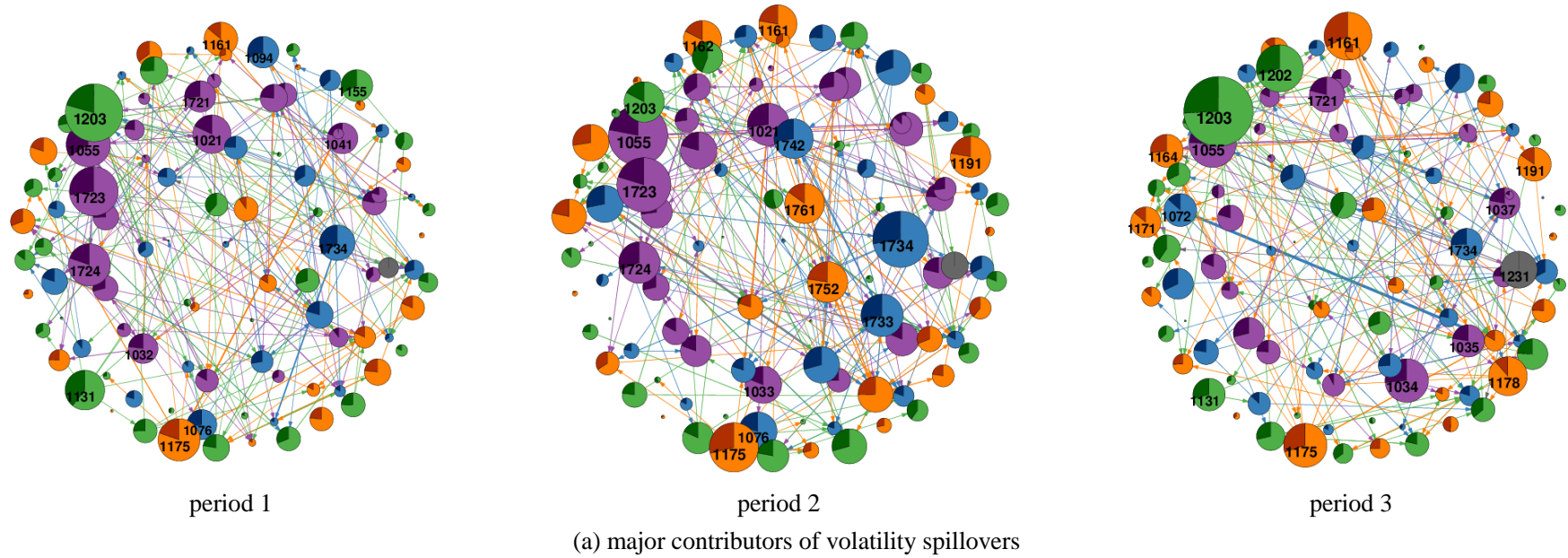

(a) major contributors of volatility spillovers

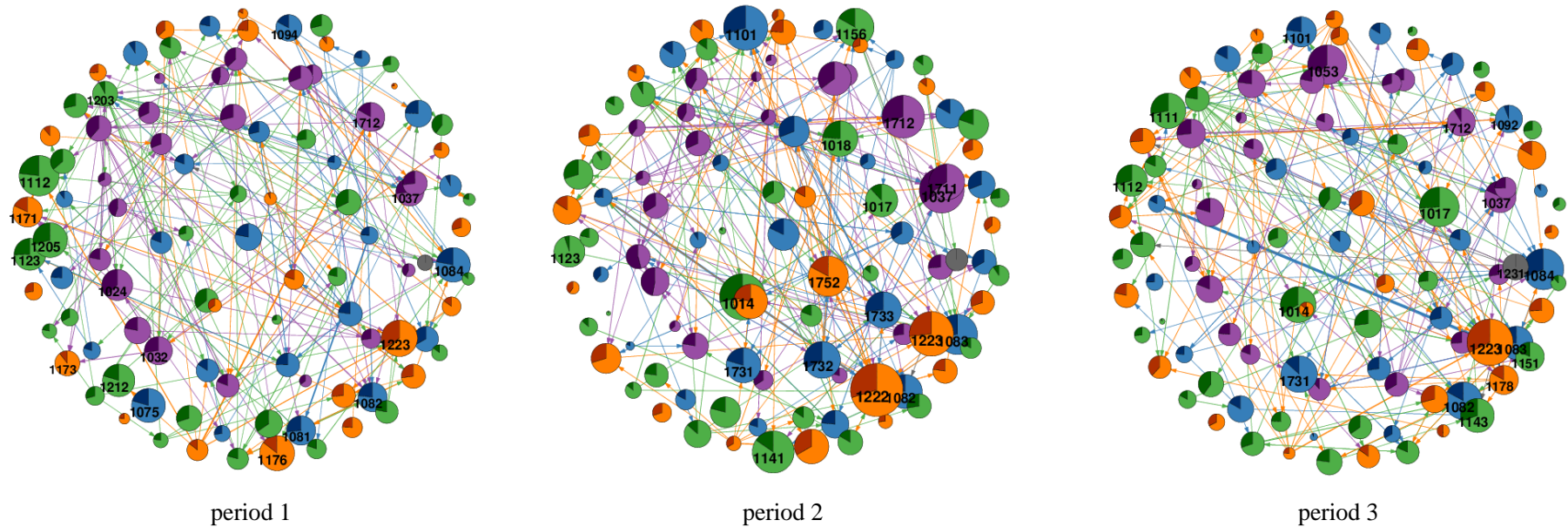

period 1     period 2     period 3

(b) major receivers of volatility spillovers

Note: 1. The node size of $v_i$ is positively correlated to $O_i$

2. The blue, green, purple and oranges nodes represent vertices of industry indices belonging to the sector group Ke, Cg, Kg and Us, respectively. The conglomerates industry securities index (801231) is represented by a grey node.

3. The nodes are showed as pie charts. The part with lighter color in the pie chart of $v_i$ represents $TOTO_i/O_i$ (in subfigure (a)) or $TIFO_i/I_i$ (in subfigure (b)).

4. The widths of edges are positively correlated with their intensity. The colors of edges are consistent with the color of nodes where they effluent.

5. Only the top 5% edges with the highest intensity of volatility spillovers are showed in each subfigure.

6. The industry securities index name corresponding with each of the 4-digit codes in the figure can be found in appendix.

**Figure 5: Major spillover effect contributors and receivers in different periods (classified by sector group)**

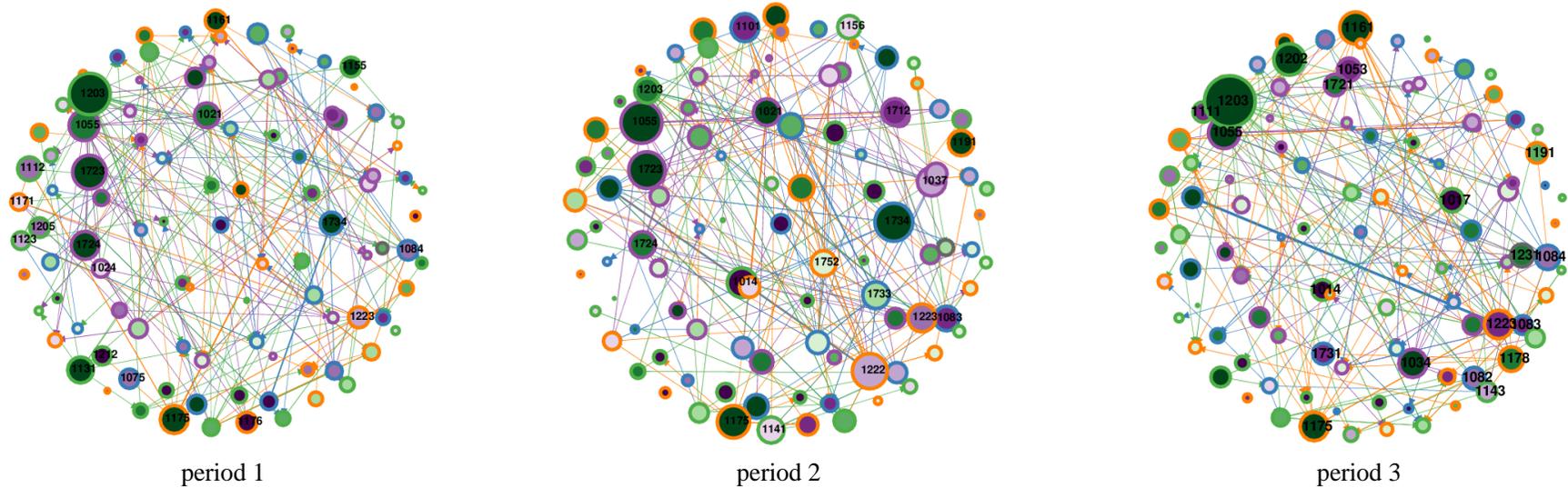

<div style="text-align:center">period 1          period 2          period 3</div>

Note:1.The node size of $v_i$ is positively correlated to $\max\{I_i, O_i\}$

2. The node color of $v_i$ with high $ri_i$ is close to dark green. In contrast, the color of the one with low $ri_i$ is close to purple.

3. The nodes with blue, green, purple and oranges frames represent vertices of industry indices belonging to the sector group Ke, Cg, Kg and Us, respectively. The conglomerates industry securities index (801231) is represented by a node with grey frame.

4. The widths of edges are positively correlated with their intensity. The colors of edges are consistent with the color of the frames of the nodes where they effluent.

5. Only the top 5% edges with the highest intensity of volatility spillovers are showed in each subfigure.

6. The industry securities index name corresponding with each of the 4-digit codes in the figure can be found in appendix.

**Figure 6: Gross and net spillover effect of the nodes in spillover networks in different periods**

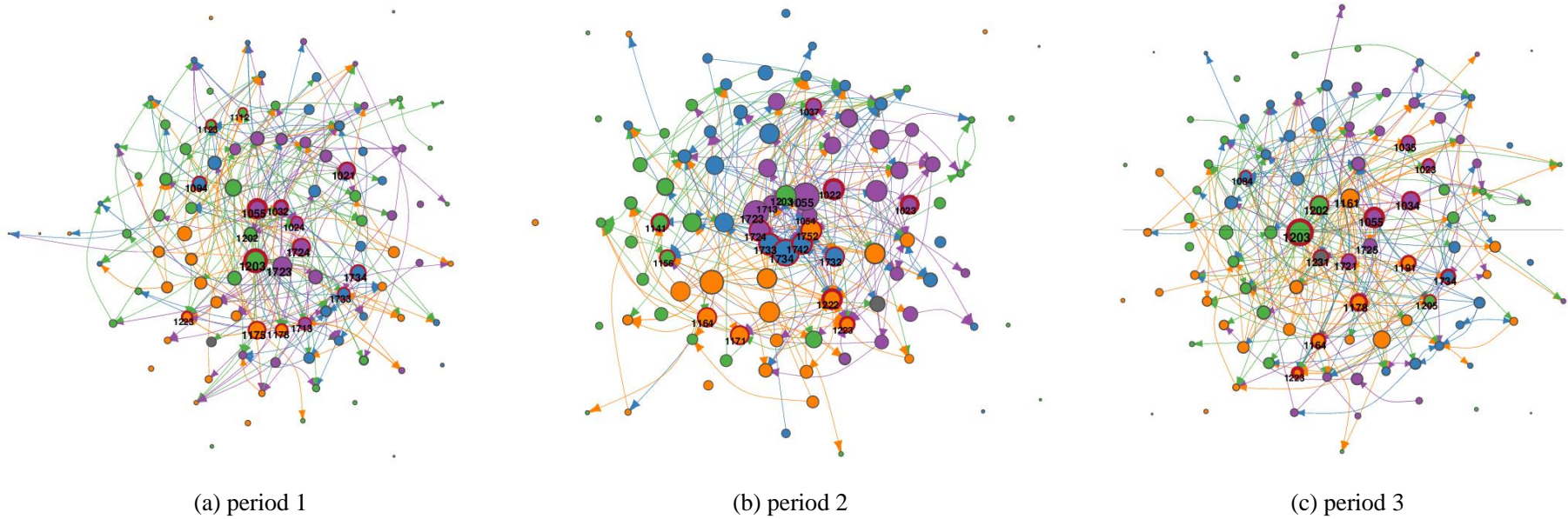

(a) period 1            (b) period 2            (c) period 3

Note:1. The node size of $v_i$ is positively correlated to $O_i$

2. The blue, green, purple and oranges nodes represent vertices of industry indices belonging to the sector group Ke, Cg, Kg and Us, respectively. The conglomerates industry securities index (801231) is represented by a grey node.

3. The nodes that make a ring in the middle of the picture are those with the highest k-shell level in each network.

4. The nodes with red ring represent the top 15 industries with the highest $WBC$ in each spillover network

5. The industry securities index name corresponding with each of the 4-digit codes in the figure can be found in appendix.

**Figure 7: The centrality and k-shell decomposition structure of the nodes in different periods**